\documentclass[letterpaper]{article}
\usepackage[preprint]{aaai2027}
\usepackage[hyphens]{url}
\usepackage{graphicx}
\usepackage{natbib}
\usepackage{caption}
\usepackage{algorithm}
\usepackage{algorithmic}
\usepackage{newfloat}
\usepackage{listings}
\DeclareCaptionStyle{ruled}{labelfont=normalfont,labelsep=colon,strut=off}
\floatstyle{ruled}
\newfloat{listing}{tb}{lst}{}
\floatname{listing}{Listing}
\usepackage{booktabs}
\usepackage{amsmath}
\usepackage{amssymb}
\usepackage{array}
\DeclareUnicodeCharacter{2212}{\ensuremath{-}}

\graphicspath{{Figures/}{supplementary_figures/}}
\title{AcoustiTrace: When Plausible Sound Violates Physics}
\author{
    Shiyang Li\textsuperscript{\rm 1,\rm 2},
    Yuewen Cao\textsuperscript{\rm 2},
    Yihao Liu\textsuperscript{\rm 2},
    Yuandong Pu\textsuperscript{\rm 3},\\
    Baochang Zhang\textsuperscript{\rm 4},
    Xiaofei Li\textsuperscript{\rm 5},
    Changqing Zou\textsuperscript{\rm 1,\rm 6}
}
\affiliations{
    \textsuperscript{\rm 1}Zhejiang University\\
    \textsuperscript{\rm 2}Shanghai Artificial Intelligence Laboratory\\
    \textsuperscript{\rm 3}Shanghai Jiao Tong University\\
    \textsuperscript{\rm 4}Beihang University\\
    \textsuperscript{\rm 5}School of Engineering, Westlake University\\
    \textsuperscript{\rm 6}Zhejiang Lab\\
    12521030@zju.edu.cn
}
\copyrighttext{Project page: \url{https://leader-sheng.github.io/AcoustiTrace/}}

\begin{document}
\maketitle

\begin{abstract}
  Recent audio--video generators can produce semantically plausible and apparently synchronized sound, yet may still violate the acoustic processes implied by visible events and environments. Existing benchmarks provide limited support for attributing such violations to particular acoustic processes and quantifying their severity. We introduce AcoustiTrace, a diagnostic benchmark that formalizes acoustic physical realism in audio--video generation. AcoustiTrace organizes text-to-audio-video (T2AV) and image-to-audio-video (I2AV) evaluation around the acoustic process, covering sound generation, propagation environment, and acoustic reception through eight dimensions grounded in measurable acoustic quantities. Based on these evaluation dimensions, we construct a large-scale dataset organized around acoustic mechanisms, comprising real-world audio--video recordings and acoustically annotated RGB-D observations, and use it to develop targeted prompt suites and validated evaluators. Experiments reveal that even leading generators still struggle with fundamental acoustic processes despite producing plausible sound events. Finally, we show that the diagnostics AcoustiTrace provides for specific acoustic relations can guide model refinement toward more physically faithful audio and open new directions for incorporating acoustic principles into training objectives, reward modeling, and candidate selection.
\end{abstract}

\begin{figure}[!t]
  \centering
  \includegraphics[width=0.9\columnwidth]{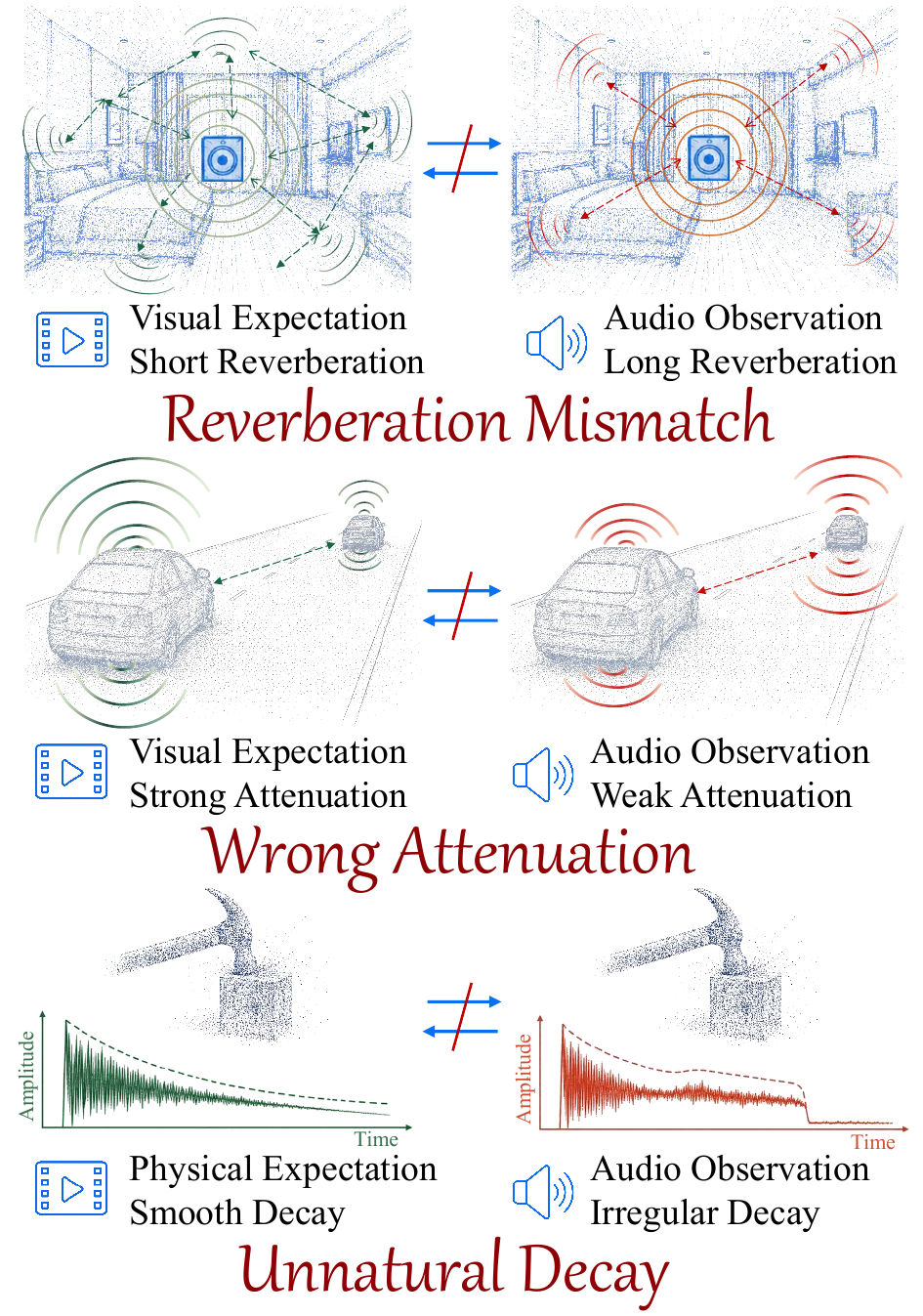}
  \caption{Representative cases where plausible sound violates acoustic physics.
  AcoustiTrace systematically diagnoses these and additional acoustic
  failures across eight evaluation dimensions.
  }
  \label{fig:motivation}
\end{figure}

\begin{figure*}[t]
  \centering
  \includegraphics[width=0.90\textwidth]{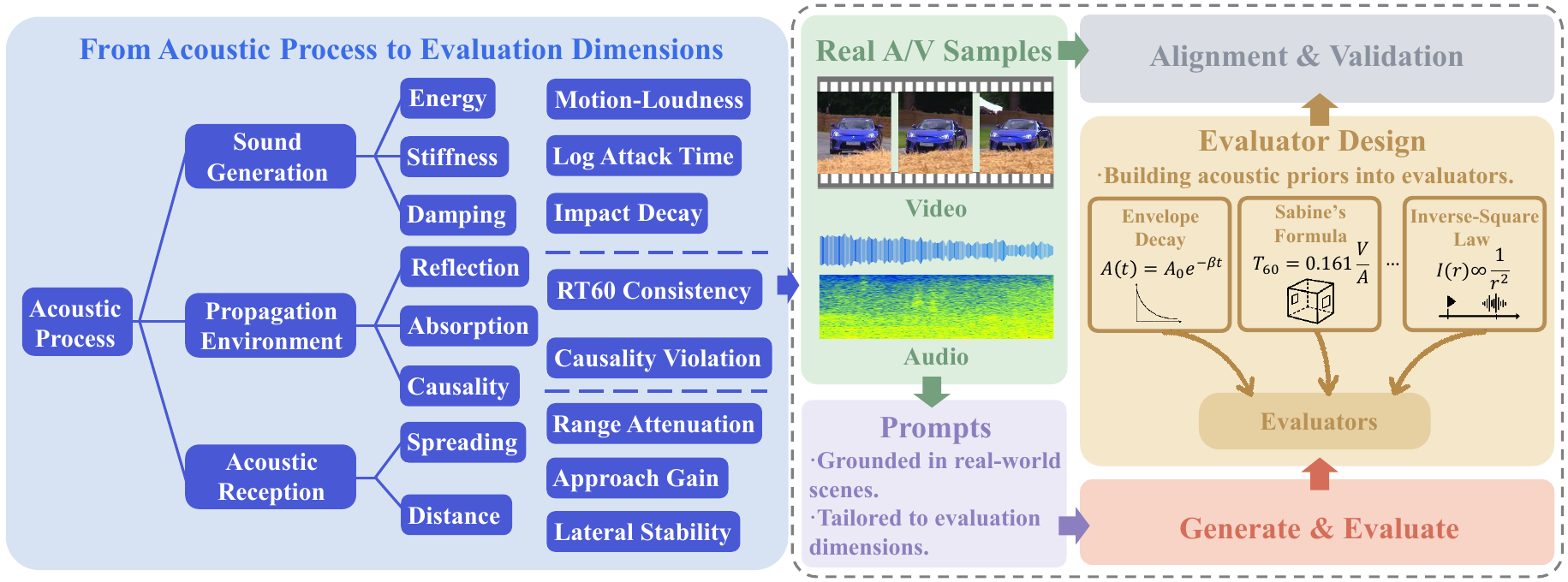}
  \caption{Framework of AcoustiTrace. The benchmark organizes acoustic physical
realism into sound generation, propagation environment, and acoustic
reception across eight evaluation dimensions. Real-world A/V data
support prompt construction, evaluator design, and validation. By
embedding acoustic priors into specialized evaluators, AcoustiTrace
supports systematic diagnosis of acoustic physical realism in joint A/V
generation.}
  \label{fig:framework}
\end{figure*}

\section{Introduction}
\label{sec:introduction}

A growing class of generative models jointly synthesizes video and
audio
\citep{liu2024syncflow,liu2025javisdit,wang2025universe,
low2025ovi,hacohen2026ltx2}. Benchmark studies indicate that these
approaches have made substantial progress in cross-modal semantic coherence
and temporal synchronization
\citep{mao2024tavgbench,hua2026vabench}. Nevertheless, perceptual plausibility and audio--video alignment do not
guarantee fidelity to the underlying acoustic processes
\citep{li2026flatsounds,oh2026pavas,cui2026avphys,
xie2025phyavbench}. Such violations may include unnatural attack or decay, mismatched
reverberation, or incorrect distance attenuation. They may occur even
when the expected sound is present and apparently synchronized. This is
precisely when plausible sound violates physics.

To assess such failures, recent work has expanded audio--video
evaluation to explicitly consider physical realism \citep{xie2025phyavbench,li2026flatsounds,cui2026avphys}. Yet a diagnostic gap remains: existing approaches do not systematically combine validated and interpretable acoustic measurements to diagnose generative models. Consequently, they provide limited support for localizing failures within the acoustic process, quantifying the magnitude of departures from expected acoustic relations, and translating these diagnoses into model refinement.

To address this gap, we introduce \textbf{AcoustiTrace}, a diagnostic benchmark that formalizes acoustic physical realism in terms of the classical sequence of sound generation, propagation through the environment, and reception by the listener
\citep{kinsler2000fundamentals}. As illustrated in
Figure~\ref{fig:framework}, AcoustiTrace translates this decomposition
into eight evaluation dimensions for text-to-audio-video (T2AV) and
image-to-audio-video (I2AV) generation. Each dimension tests an explicit
acoustic relation between observable visual evidence and a measurable
audio quantity. The resulting diagnostics are expressed in interpretable
acoustic terms using established descriptors of temporal envelopes,
reverberation decay, and acoustic propagation
\citep{peeters2011timbre,schroeder1965reverberation,
kinsler2000fundamentals}.

Guided by these dimensions, we curate a dataset organized around
real-world acoustic processes and use it to develop the evaluators,
which we validate with real-world and ground-truth anchors together with
controlled perturbations. The dataset then informs 605-prompt T2AV and
748-prompt I2AV suites targeting acoustic relations within the
evaluators' validated measurement scope. Using AcoustiTrace, we evaluate nine joint audio--video generators. Experiments reveal that even leading generators
still struggle to faithfully model acoustic relations across the acoustic process despite
producing plausible sound events. This persistent gap highlights an
urgent need to improve current models through acoustically cleaner
training subsets, explicit acoustic supervision, and specialized
annotations targeting acoustic quantities.

To explore one such direction, we conduct an initial intervention study
by converting a diagnosed range attenuation error into a guidance signal
for audio sampling. With the model parameters, prompt, and video stream
held fixed, the guidance improves consistency with the targeted acoustic
relation in 80.16\% of valid samples. This proof of concept demonstrates
that the validated and interpretable diagnostics provided by
AcoustiTrace can identify concrete targets and guide model refinement,
opening broader opportunities for physics-aware training objectives,
reward modeling, and candidate selection.

Our contributions are summarized as follows:
\begin{enumerate}

  \item We introduce AcoustiTrace, a diagnostic benchmark that organizes
  the evaluation of joint audio--video generation around the acoustic
  process. Across eight evaluation dimensions, AcoustiTrace uses validated
  and interpretable acoustic measurements to identify and quantify
  deviations from expected acoustic relations. 

  \item We curate a dataset centered on acoustic processes, comprising
  11,296 unique real-world audio--video samples and 82,828 RGB-D
  samples with acoustic absorption maps and RT60 labels. The dataset supports evaluator development and validation and informs
  the construction of 605-prompt T2AV and 748-prompt I2AV suites, enabling
  systematic evaluation across diverse acoustic conditions.

  \item We demonstrate that AcoustiTrace diagnostics identify concrete
  targets for model refinement. A targeted intervention based on a
  diagnosed range attenuation error shows how one such diagnosis can be
  converted into guidance for audio sampling, pointing to broader
  opportunities for incorporating acoustic principles into training
  objectives, reward modeling, and candidate selection.

\end{enumerate}

\section{Related Work}
\subsection{Audio--Video Generation Models}
Audio generation for video follows two main directions. Video-to-audio
methods synthesize a soundtrack for an existing video, with recent
diffusion- and flow-based systems improving semantic relevance, temporal
alignment, and audio quality
\citep{zhang2024foleycrafter,cheng2025mmaudio,
shan2025hunyuanvideofoley,jeong2025rewas}. Joint audio--video models instead generate
both modalities in a shared process, using temporal flow alignment,
expert stitching, parallel backbones, or unified multimodal
transformers
\citep{liu2024syncflow,wang2025universe,low2025ovi,
hacohen2026ltx2,openmoss2026mova,ji2026nava,
liu2026javisditpp}. These systems primarily optimize perceptual quality,
semantic coherence, and synchronization; their fidelity to the acoustic
process implied by the scene remains less systematically characterized.
\subsection{Evaluation of Video and Audio--Video Generation}
Evaluation has progressed from multidimensional video benchmarks such as
VBench to audio--video benchmarks and metrics for perceptual quality, semantic
alignment, synchronization, spatial correspondence, and instruction
following
\citep{huang2024vbench,yariv2024diverse,shimada2024savgbench,
hua2026vabench,cao2025t2avcompass}. Recent work further targets physical
realism: PhyAVBench measures responses to controlled acoustic changes,
FlatSounds tests single-factor physical trends, PAVAS relates estimated
impact physics to acoustic attributes, and AV-Phys Bench evaluates
physical consistency within and across modalities
\citep{xie2025phyavbench,li2026flatsounds,
oh2026pavas,cui2026avphys}. Together, these studies establish important paradigms for evaluating
physical sensitivity, trend compliance, and physical commonsense.
AcoustiTrace complements these efforts by organizing validated acoustic measurements around the acoustic process. It
localizes which acoustic relation fails and quantifies the corresponding
deviation in interpretable physical terms.

\section{AcoustiTrace}
\label{sec:method}

\subsection{Benchmark Design}
\label{sec:benchmark_design}

AcoustiTrace evaluates acoustic physical realism in generated
audio--video content by examining whether the generated soundtrack is
consistent with visible events and scene conditions. Its test cases are
organized around the acoustic process, which comprises three stages:
sound generation, propagation environment, and
acoustic reception.

\begin{figure}[t]
  \centering
  \includegraphics[width=0.9\columnwidth]{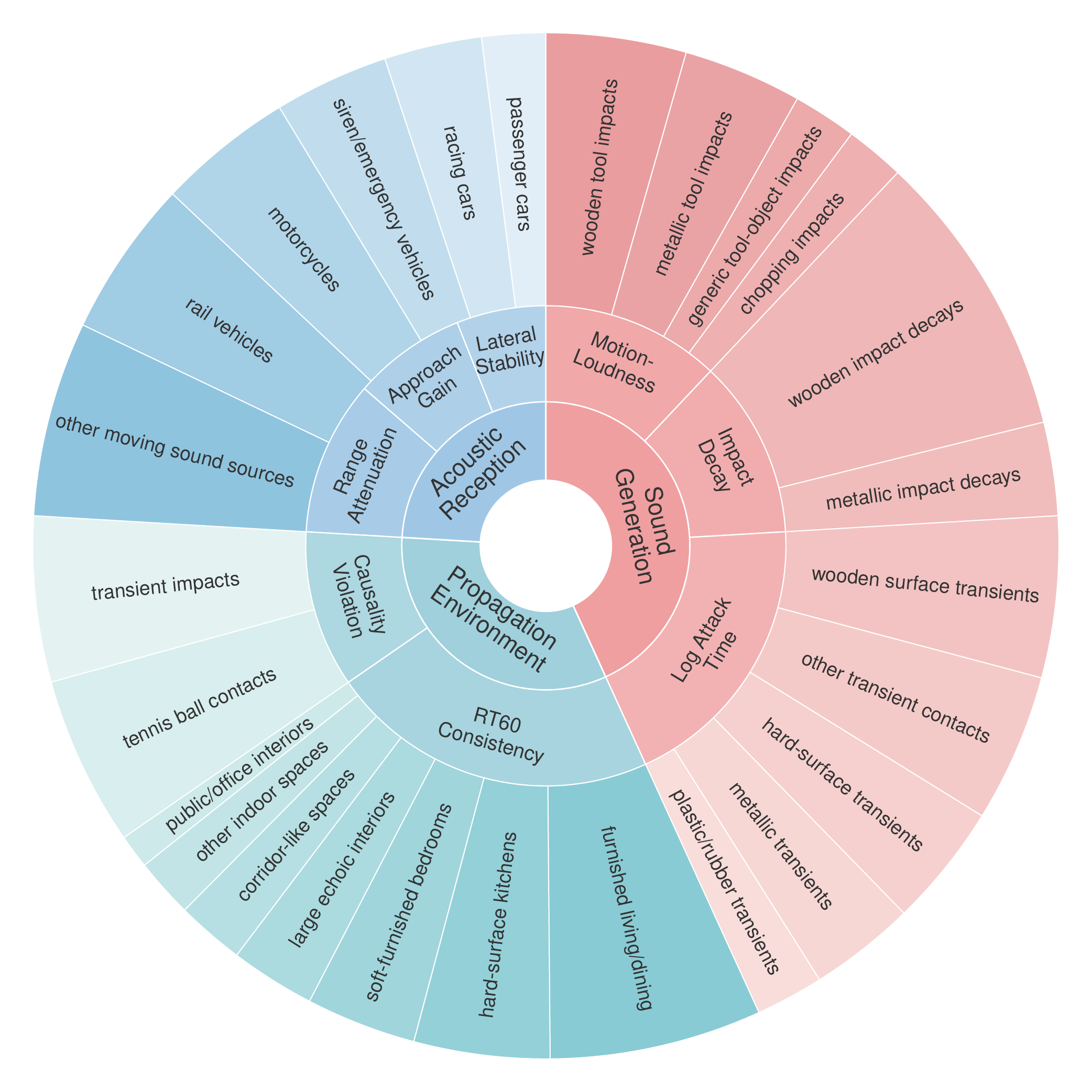}
  \caption{
    Distribution of the AcoustiTrace prompt suites. The sunburst chart shows
    their hierarchical composition, with sector area proportional to prompt
    count.
    }
  \label{fig:prompt_distribution}
  \end{figure}

  This organization yields eight evaluation dimensions:
  Motion--Loudness, Log Attack Time, Impact Decay, RT60 Consistency,
  Causality Violation, Range Attenuation, Approach Gain, and Lateral
  Stability. Figure~\ref{fig:prompt_distribution} summarizes the prompt
  suites, which contain 605 T2AV and 748 I2AV prompts.
  
  As shown in Figure~\ref{fig:framework}, AcoustiTrace links data
  curation, evaluator validation, and prompt construction: the same
  acoustic relations guide dataset construction, evaluator design, and
  prompt-suite coverage. This keeps the benchmark scope aligned with
  validated evaluator capabilities.

\subsection{Dataset Construction}
\label{sec:data_construction}

Guided by the eight evaluation dimensions, we construct a dataset that
combines real-world audio--video anchors with a dedicated RGB-D
collection containing Acoustic Alpha Maps and RT60 labels. The dataset
supports evaluator development, validation, and prompt design. Further
details are provided in the supplementary material.

\paragraph{Real-world audio--video anchors}
We retrieve candidate videos from public sources using queries tailored
to observable acoustic relations, such as different motion strengths,
isolated impacts, timed contacts, approaching or receding sources, and
motion at approximately constant range. Candidates then pass automated
multimodal large language model (MLLM) and task-specific detector filters, followed by human review to
confirm that the target relation is observable and measurable in both
modalities. This process yields 11,296 unique real-world audio--video
clips.

\paragraph{Acoustically annotated RGB-D observations}
RT60 Consistency assesses whether the reverberation implied by the
visible environment agrees with the apparent RT60 estimated from
generated audio. Its visual representation is guided by the Sabine
relation, which links reverberation time to room volume and equivalent
absorption area \citep{kinsler2000fundamentals}:
\[
    T_{60}
    =
    0.161\frac{V}{A},
    \qquad
    A
    =
    \sum_j S_j\alpha_j,
\]
where \(V\) denotes room volume, \(S_j\) is the area of boundary region \(j\), and \(\alpha_j\) is its frequency-dependent absorption
coefficient. Under this relation, visual estimation of RT60 requires
geometric cues related to \(V\) and \(S_j\), together with spatially
resolved absorption cues related to \(\alpha_j\).

To our knowledge, no existing large-scale dataset directly pairs indoor
RGB-D observations with pixel-level surface absorption maps and
band-specific RT60 labels. We therefore construct 82,828 samples, each
combining a Matterport3D RGB-D observation, a PTB-derived Acoustic Alpha
Map, and a 500 Hz RT60 label estimated from a SoundSpaces~2.0 room
impulse response
\citep{chang2017matterport3d,ptb_absorption_database,chen2022soundspaces2,schroeder1965reverberation}.

\begin{figure}[!t]
  \centering
  \includegraphics[width=\columnwidth]{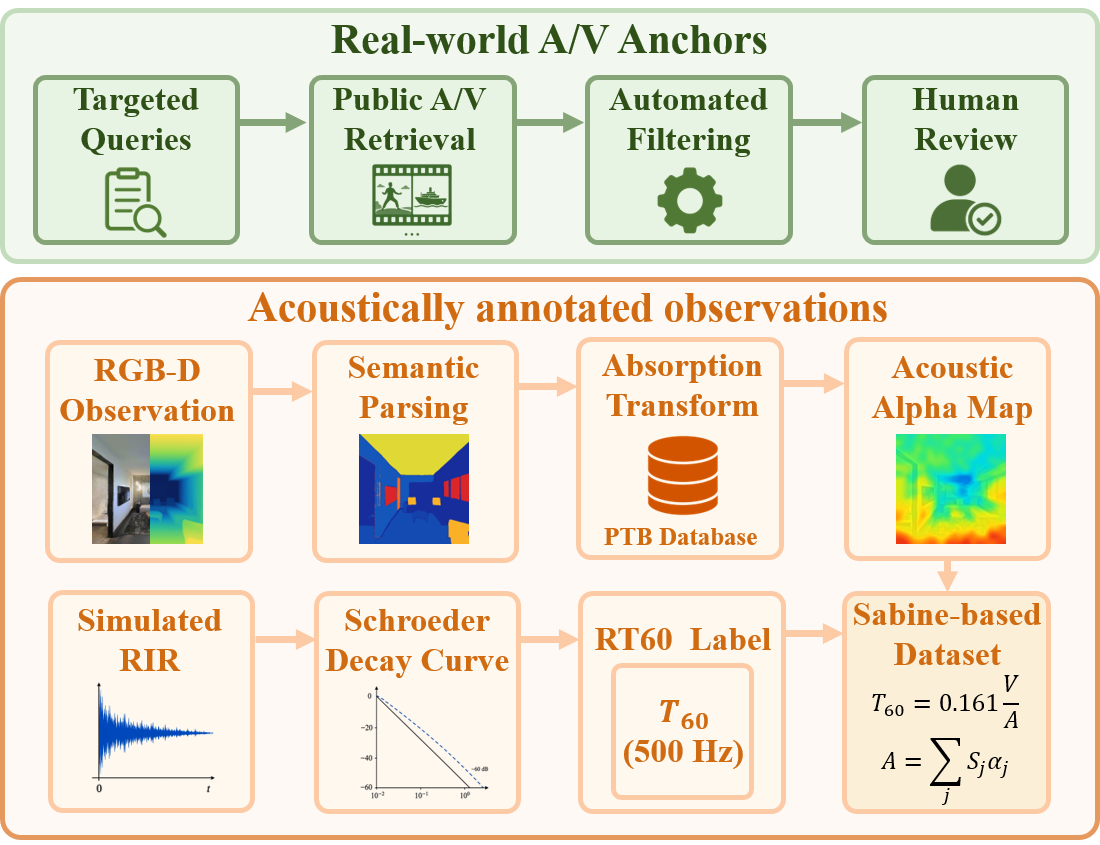}
  \caption{Construction of the AcoustiTrace dataset from real-world
  audio--video anchors and acoustically annotated RGB-D observations.}
  \label{fig:dataset_construction}
\end{figure}

\subsection{Acoustic Evaluation Dimensions}
\label{sec:evaluators}

AcoustiTrace evaluates eight acoustic relations spanning sound
generation, propagation environment, and acoustic reception. Each
dimension specifies the required visual evidence, audio measurement, and
expected acoustic relation. Each evaluator is applied only when its
required visual evidence and corresponding audio event or segment can be reliably localized and measured. Dimension-specific validity criteria, detector settings, and score mappings are provided in the supplementary material.

\subsubsection{Sound Generation: Motion--Loudness}
\label{sec:motion_loudness}

Motion--Loudness probes the coupling between visible excitation
strength and emitted acoustic energy. A stronger action is expected to produce a louder event. The evaluator tests whether this
relation is preserved between actions within the same generated sample. The evaluation first employs two expert models for event localization in the visual and audio streams \citep{zhou2025ovavel,hai2025flexsed}. The localized actions are then compared by an MLLM using visual evidence only. The MLLM ranks their visible action strength according to motion amplitude, speed, motion range, and duration. The evaluator checks whether the visually stronger action is paired with the higher audio level.

\subsubsection{Sound Generation: Log Attack Time}
\label{sec:lat}

Log Attack Time evaluates whether the onset of a generated impact is
consistent with the material and contact conditions of a real reference.
For each I2AV test case, the conditioning image and reference audio are
drawn from the same temporally aligned Greatest Hits recording, in which
a drumstick strikes a visible object \citep{owens2016visually}. Hard and
rigid contacts generally exhibit faster onsets than soft or compliant
contacts. We localize the corresponding impacts in the generated and
reference audio and compute Log Attack Time for each; a smaller absolute
difference indicates closer onset agreement. This dimension is restricted
to I2AV because the conditioning image identifies the struck object and
its material context.

\subsubsection{Sound Generation: Impact Decay}
\label{sec:impact_decay}
Impact Decay evaluates whether the sound following an isolated visible
impact exhibits a physically plausible loss of energy. An impact excites
transient structural vibrations that dissipate through material damping
and acoustic radiation, often producing an approximately exponential
decay. The evaluator measures how well the post-impact energy envelope
follows this general dissipative pattern using the coefficient of
determination of an exponential fit. Higher fit quality indicates more plausible impact
decay.

\subsubsection{Propagation Environment: RT60 Consistency}
\label{sec:rt60_consistency}

RT60 Consistency evaluates whether the reverberation present in the
generated audio agrees with that implied by the visible environment.
Specifically, this dimension compares the RT60 inferred from visual
scene cues with the apparent RT60 estimated from the generated audio. On the visual side, we fine-tune an MLLM equipped with a Sabine-guided
physics head to estimate RT60 from the visible scene. Details of the MLLM fine-tuning strategy and the design of
the physics head are provided in the supplementary material. On the
audio side, we estimate the apparent RT60 from a noise-compensated
Schroeder energy decay curve.

\subsubsection{Propagation Environment: Causality Violation}
\label{sec:causality_violation}

Causality Violation evaluates whether an audible event begins before
the visible event that produces it. Using expert models, we
localize and temporally match the corresponding visual and audio events. The detected audio onset is then compared with the visible contact time. An audio onset that precedes the visible event is counted as a causality violation.

\subsubsection{Acoustic Reception: Range Attenuation}
\label{sec:range_attenuation}
Range Attenuation evaluates whether the generated audio preserves the
expected distance-dependent attenuation as a visible sound source moves
farther from the listener. In an ideal free field with approximately
constant source power, sound intensity follows the inverse-square law
\citep{kinsler2000fundamentals}. We track the dominant sound source in
the video, combine its spatial trajectory with depth estimates to obtain
a relative range curve, and align this curve with the short-time audio
level trajectory. The observed relative level change is compared with
the distance-attenuation reference derived from the inverse-square law,
and the coefficient of determination \(R^2\) measures how well this
reference explains the generated audio level trajectory.
\subsubsection{Acoustic Reception: Approach Gain}
\label{sec:approach_gain}

Approach Gain applies to intervals in which the tracked source moves
toward the listener. It evaluates whether decreasing visual range is
accompanied by a corresponding increase in received audio level. Flattened, reversed, or
temporally inconsistent gain patterns receive lower scores.

\begin{table*}[t]
\centering
\setlength{\tabcolsep}{1mm}
\begin{tabular}{@{}llccccccccc@{}}
\toprule
\multicolumn{1}{@{}l}{Task} & Evaluator & LTX-2.3 & JavisDiT++ & NAVA & Ovi & UniVerse-1 & MOVA & Veo 3.1$^{*}$ & Seedance 2.0$^{*}$ & \multicolumn{1}{c@{}}{Wan 2.7$^{*}$} \\
\midrule
\textit{T2AV} & Range Attenuation & 81.63 & 39.17 & 68.12 & 57.13 & -- & 62.42 & 71.85 & \textbf{86.48} & 74.49 \\
& Approach Gain & 74.11 & 68.39 & 69.88 & 70.62 & -- & 81.14 & \textbf{88.68} & 71.32 & 71.69 \\
& Lateral Stability & \textbf{92.18} & 87.55 & 84.42 & 89.56 & -- & 86.80 & 81.79 & 76.57 & 82.03 \\
& Motion--Loudness & \textbf{88.51} & 51.52 & 75.86 & 56.44 & -- & 74.00 & 75.10 & 83.98 & 82.37 \\
& Impact Decay & 90.00 & 73.62 & 88.59 & \textbf{98.62} & -- & 59.10 & 98.61 & 91.39 & 85.91 \\
& Causality Violation & 86.52 & 86.60 & 93.22 & 88.89 & -- & 83.33 & 98.62 & \textbf{99.16} & 95.68 \\
& RT60 Consistency & 41.08 & 69.81 & 69.74 & 69.07 & -- & 53.57 & \textbf{73.51} & 24.28 & 33.20 \\
% & Compact mean & 79.15 & 68.09 & 78.55 & 75.76 & -- & 71.48 & \textbf{85.45} & 76.17 & 75.05 \\
\midrule
\textit{I2AV} & Range Attenuation & 66.89 & -- & \textbf{67.78} & 45.30 & 32.53 & 53.50 & 63.70 & 59.98 & 45.17 \\
& Approach Gain & 65.22 & -- & 65.88 & 71.98 & 65.81 & 78.79 & \textbf{85.76} & 70.83 & 76.12 \\
& Lateral Stability & \textbf{91.33} & -- & 86.91 & 91.08 & 84.70 & 86.01 & 81.14 & 79.50 & 77.38 \\
& Motion--Loudness & 72.73 & -- & \textbf{89.55} & 54.33 & 50.77 & 80.90 & 84.53 & 83.92 & 85.67 \\
& Impact Decay & 84.95 & -- & 83.64 & 83.20 & 63.13 & 84.28 & 83.48 & \textbf{87.39} & 72.46 \\
& Causality Violation & 88.39 & -- & 87.80 & 50.60 & 44.97 & 90.00 & \textbf{97.36} & 97.03 & 95.75 \\
& Log Attack Time & 62.56 & -- & 65.29 & 50.42 & 34.65 & 63.88 & 73.28 & 50.15 & \textbf{73.70} \\
& RT60 Consistency & 42.12 & -- & 70.71 & 66.91 & 42.06 & 55.01 & 46.17 & 17.22 & \textbf{76.28} \\
% & Compact mean & 71.77 & 65.97 & \textbf{77.20} & 64.23 & 52.33 & 74.05 & 76.93 & 68.25 & 75.32 \\
\bottomrule
\end{tabular}
\caption{AcoustiTrace results for T2AV and I2AV generation.
Bold values indicate the best model result for each metric within each
task. Higher values indicate greater physical consistency.
Models marked with $^{*}$ are proprietary systems evaluated through
official interfaces. A dash indicates that the evaluated model release
does not natively support the corresponding task.}
\label{tab:combined-results}
\makeatletter
\def\@currentlabel{\thetable(a)}\label{tab:t2av-results}
\def\@currentlabel{\thetable(b)}\label{tab:i2av-results}
\makeatother
\end{table*}

\subsubsection{Acoustic Reception: Lateral Stability}
\label{sec:lateral_stability}

Lateral Stability evaluates the received level when the source remains
at approximately constant range or moves primarily across the listener's
field of view. After selecting intervals with limited variation in
relative depth, the evaluator measures systematic level drift and
short-time loudness variation. Generated audio is considered more
physically consistent when its level remains stable in the absence of a
substantial range change.

\section{Benchmarking Audio--Video Generators}

\subsection{Main Results}
\label{sec:main_results}

We evaluate nine joint audio--video generators: LTX-2.3~\citep{hacohen2026ltx2}, JavisDiT++~\citep{liu2026javisditpp},
NAVA~\citep{ji2026nava}, Ovi~\citep{low2025ovi}, UniVerse-1~\citep{wang2025universe}, and MOVA~\citep{openmoss2026mova}, along with Veo 3.1~\citep{deepmind2026veo3modelcard, deepmind2025veo3}, Seedance 2.0~\citep{seedance2026seedance2},
and Wan 2.7~\citep{alibabacloud2026wan27}. We use each model's supported settings.
Table~\ref{tab:combined-results} reports conditional scores on valid
outputs. Across the reported model--evaluator cells, validity ranges from
40.5\% to 100\% (median 99.8\%); coverage is consistently lowest for
RT60, with full cell-wise rates and exclusion criteria provided in the
supplementary material.

As shown in the T2AV block of Table~\ref{tab:combined-results}, no model dominates across all dimensions. LTX-2.3 leads Motion--Loudness and Lateral Stability, Ovi leads Impact Decay, Veo 3.1 leads Approach Gain and RT60 Consistency, and Seedance 2.0 leads Range Attenuation and Causality Violation. The I2AV results show a different pattern of dimension-wise leadership. NAVA leads Range Attenuation and Motion--Loudness, LTX-2.3 leads Lateral Stability, Veo 3.1 leads Approach Gain and Causality Violation, Seedance 2.0 leads Impact Decay, and Wan 2.7 leads Log Attack Time and RT60 Consistency.

\begin{figure*}[!t]
\centering
\includegraphics[width=0.90\textwidth]{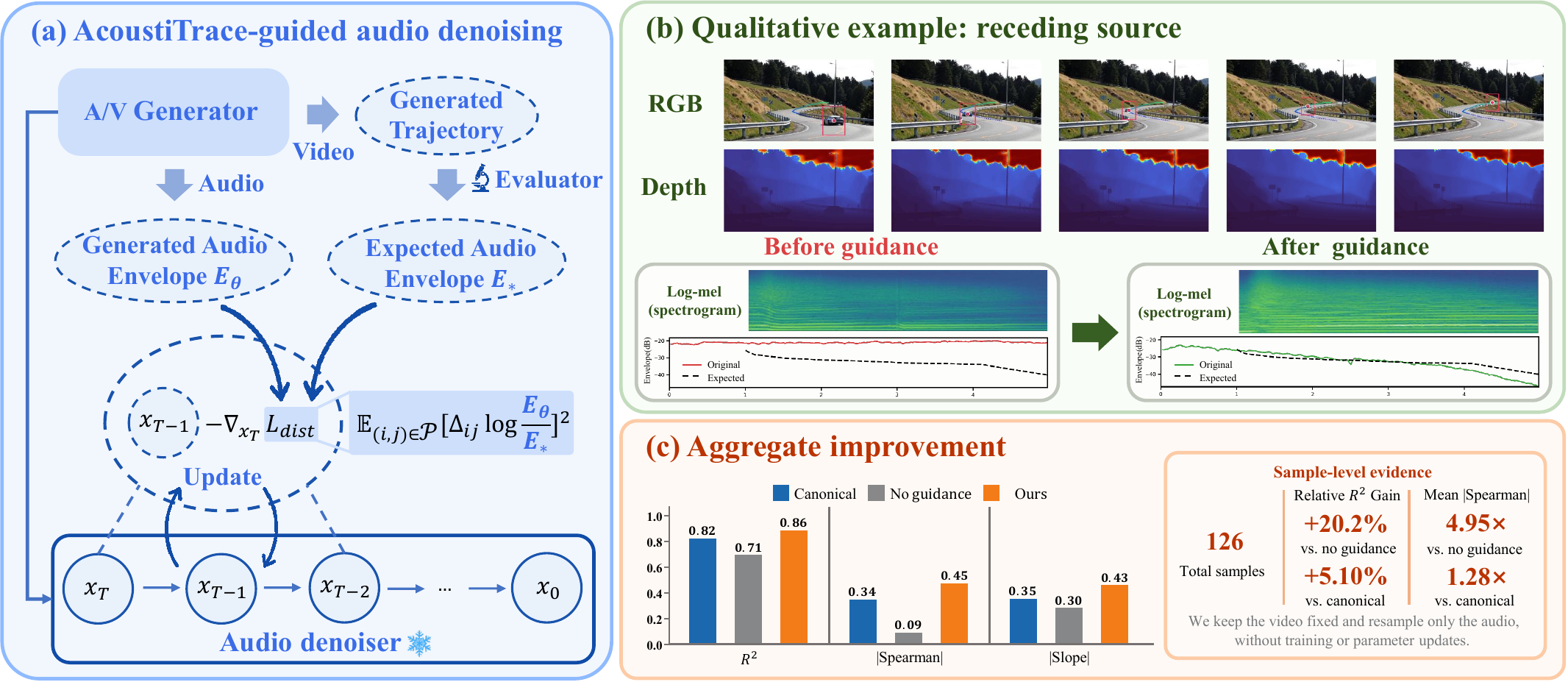}
\caption{
From diagnosis to intervention.
\textbf{(a)} The Range Attenuation residual from a fixed visual trajectory guides decoded-mel audio denoising.
\textbf{(b)} A qualitative receding-source example visualizes how guidance reshapes the audio envelope toward the expected attenuation trend.
\textbf{(c)} Aggregate results show improved Range Attenuation \(R^2\) and stronger inverse range--loudness coupling over unguided audio resampling.
}
\label{fig:guidance}
\end{figure*}

\paragraph{Mechanism-level patterns and design implications}

Across T2AV and I2AV, the results reveal a consistent divide between
acoustic relations aligned with current design priorities and those
requiring sustained modeling of geometric or environmental structure.
Although implemented through different mechanisms, including temporal
alignment, multimodal fusion, joint latent generation, and the
composition of separate modality representations, current generators
largely prioritize semantic and temporal alignment. This shared emphasis
may help explain why most models achieve high consistency scores under
the Causality Violation evaluator.

Acoustic relations that are locally explicit or qualitatively observable
within short training clips, such as the relation between visible action
magnitude and sound level, decay following an isolated impact, loudness
stability at approximately constant range, and the loudness increase
during approach, may be easier to learn than relations that depend on
spatial or environmental consistency over longer intervals. This may
explain the strong Motion--Loudness and Impact Decay performance of
several models, together with the generally robust Lateral Stability and
Approach Gain results.

By contrast, Log Attack Time, RT60 Consistency, and Range Attenuation
remain particularly challenging. Log Attack Time remains lower than
Impact Decay for eight of the nine models. This gap likely reflects two
limitations in current training and evaluation practice. First, general
A/V training data rarely provide explicit supervision linking material
and contact conditions to detailed sound onset dynamics. Second,
commonly used semantic alignment and synchronization metrics, including
embedding measures such as ImageBind
\citep{girdhar2023imagebind}, emphasize overall correspondence but are
less sensitive to local envelope structure. Because these metrics shape
model selection and optimization priorities, current systems receive
limited pressure to improve such detailed acoustic properties.
Consequently, models may learn that an impact sound should occur without
reliably learning how quickly it should rise.

For RT60 Consistency, learning physically grounded scene acoustics
requires training examples with observable reverberation decay,
alongside reliable information on room geometry and materials. Because
such clean and informative data are scarce, current models may not
reliably capture how visible room properties shape reverberation.

Range Attenuation reveals a related but distinct capability gap. The
comparatively strong Approach Gain results suggest that current models
have learned a coarse association between source distance and loudness,
but this implicit relation does not consistently extend to the
attenuation expected as distance changes. This gap likely reflects
limited supervision over source trajectories, emission strength, and
corresponding loudness curves.

Taken together, these results indicate that producing plausible sound
events does not necessarily imply faithful modeling of the underlying
acoustic mechanisms. These findings highlight an urgent need for curated training
data annotated with relevant physical quantities, together with
objectives that explicitly supervise the relations among them.

\begin{figure*}[!t]
  \centering
  \includegraphics[width=0.98\textwidth]{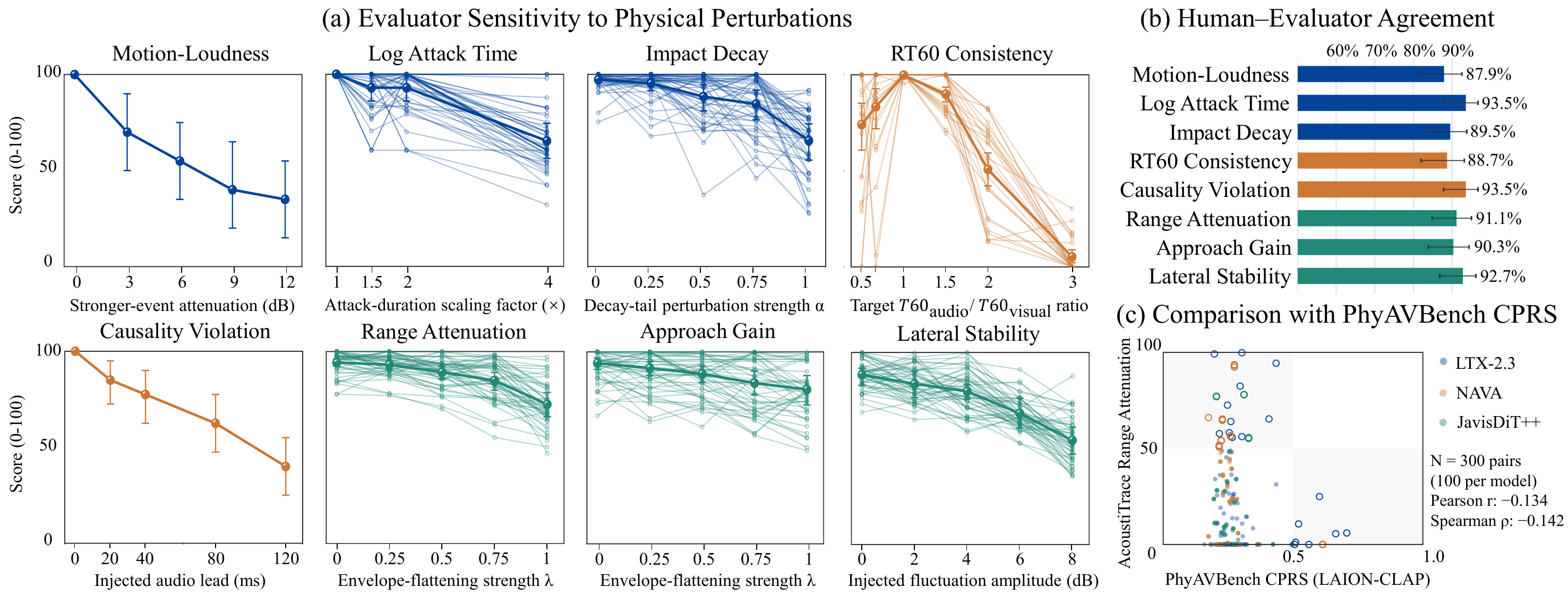}
  \caption{
    \textbf{(a)} Responses of the eight evaluators to controlled audio
    perturbations that selectively violate their targeted acoustic relations.
    The unperturbed point in each sweep corresponds to the original,
    unmodified real-world A/V samples. For RT60 Consistency, perturbation
    severity increases as the audio-to-visual RT60 ratio deviates from 1.
    \textbf{(b)} Alignment between evaluator preferences and human judgments
    on paired generated A/V samples. Error bars denote 95\% confidence
    intervals.
\textbf{(c)} AcoustiTrace Range Attenuation scores versus PhyAVBench
CPRS using LAION-CLAP on 300 pairs from
LTX-2.3, NAVA, and JavisDiT++. 
    }
  \label{fig:human_ab}
\end{figure*}

\subsection{Range-Guided Audio Sampling}

Building on these insights, we conduct an initial study to test whether AcoustiTrace diagnostics can directly guide model refinement. We target the identified Range Attenuation failure and formulate the expected attenuation relation used by the evaluator as a differentiable guidance objective for audio sampling.

Our controlled intervention on LTX-2.3 holds the generated video fixed
and resamples only the audio stream under baseline and guided conditions. From the
fixed video, we estimate source ranges \(d_t\) and guide a differentiable
decoded-mel amplitude proxy \(e_t\) via
\[
\mathcal{L}_{\mathrm{dist}}
=
\sum_{i,j}
\left[
\log\frac{e_i+\epsilon}{e_j+\epsilon}
-
\gamma\log\frac{d_j+\epsilon}{d_i+\epsilon}
\right]^2.
\]
We set \(\gamma=1\) to target inverse-distance amplitude attenuation. A
held-out calibration confirmed that the proxy preserves injected \(1/r\)
waveform scaling (\(\hat{\gamma}=0.997\)). The
objective encourages relative audio levels to follow the visual range
trajectory. Further implementation and
calibration details are provided in the supplementary material.

\begin{table}[t]
  \centering
  \begin{tabular*}{\columnwidth}{@{\extracolsep{\fill}}lccc@{}}
  \toprule
  Metric & Canonical & No guidance & Guided \\
  \midrule
  Mean $R^2$              & 0.8163 & 0.7138 & 0.8580 \\
  Spearman \(\rho\)                & $-0.3411$ & $-0.0880$ & $-0.4356$ \\
  CLAP                    & 0.1029 & 0.0498 & 0.1049 \\
  LUFS                    & $-23.601$ & $-23.039$ & $-23.586$ \\
  Audiobox PQ             & 6.031 & 6.106 & 5.884 \\
  \bottomrule
  \end{tabular*}
  \caption{Range-guided audio sampling evaluated using target attenuation metrics
  and non-target quality metrics, with the canonical video stream held
  fixed. No-guidance and guided conditions resample only the audio.}
  \label{tab:guidance}
  \end{table}

  As summarized in Table~\ref{tab:guidance} and Figure~\ref{fig:guidance}(c), audio-only resampling without guidance reduces the mean Range Attenuation \(R^2\) from 0.8163 for the canonical outputs to 0.7138. This sensitivity to audio resampling reveals instability in the default sampling process, which does not reliably reproduce the distance-dependent attenuation expected from acoustic propagation.
  
  Distance guidance increases the mean Range Attenuation \(R^2\) from 0.7138 to 0.8580 and outperforms unguided resampling in 80.16\% of cases. The mean Spearman correlation shifts from −0.0880 to −0.4356, indicating a stronger inverse relation between visual range and audio level. The qualitative example in Figure~\ref{fig:guidance}(b) shows that guidance yields an envelope that more closely follows the expected attenuation trajectory.
  
We assess non-target effects using CLAP similarity~\citep{elizalde2023clap}, integrated LUFS~\citep{itu2023bs1770,ebu2023r128}, and the Production Quality (PQ) score from Meta Audiobox Aesthetics
\citep{tjandra2025audioboxaesthetics} on the same fixed-video pairs. Distance guidance increases CLAP similarity from 0.0498 to 0.1049. Integrated loudness remains nearly identical to the canonical outputs. Audiobox Aesthetics PQ decreases slightly but remains close to the canonical score of 6.0306. Overall, the intervention improves the targeted attenuation relation and text--audio alignment without relying on global loudness adjustment, while largely preserving predicted production quality.

This intervention demonstrates a concrete path from diagnostic measurement to model refinement. AcoustiTrace identifies a failed acoustic relation, quantifies the corresponding deviation, and converts the diagnosis into an executable objective for generation. More broadly, the acoustic relations revealed by our benchmark can inform future training objectives, reward models, and candidate-selection strategies.

\subsection{Evaluator Validation and Human Alignment}

We assess the reliability of AcoustiTrace through complementary validation experiments covering real-world relation recovery, controlled perturbation sensitivity, and human alignment on generated A/V content. We first test all evaluators on our collected
real-world A/V dataset to determine whether they recover the expected
acoustic relations, with the original samples represented by the
unperturbed points in Figure~\ref{fig:human_ab}(a). We then apply
controlled audio perturbations to selectively violate the acoustic
relation targeted by each evaluator. As shown in
Figure~\ref{fig:human_ab}(a), all evaluators respond in the expected
direction as the targeted violation becomes more severe.

We also compare evaluator preferences with human judgments. A panel of
30 expert raters provides more than 1,000 judgments on paired generated
A/V samples, indicating which sample better satisfies the target
acoustic relation. The high agreement shown in
Figure~\ref{fig:human_ab}(b) indicates strong alignment between evaluator
preferences and human judgments. Detailed validation protocols
and quantitative results are provided in the supplementary material.

\subsection{Comparison with PhyAVBench CPRS}

We further compare AcoustiTrace with PhyAVBench's official LAION-CLAP
CPRS on its inverse-square-law near--far prompt pairs. Across 300 valid
pairs generated by LTX-2.3, NAVA, and JavisDiT++, CPRS
and the AcoustiTrace Range Attenuation score exhibit only weak
correlation. CPRS
measures whether the near-to-far embedding transition follows the
official reference direction, whereas AcoustiTrace evaluates whether the
visually estimated distance change between the near and far samples is
accompanied by the corresponding audio attenuation. As shown in
Figure~\ref{fig:human_ab}(c), the results include both
high-CPRS/low-AcoustiTrace and low-CPRS/high-AcoustiTrace cases,
indicating that embedding-transition sensitivity and relation-specific
acoustic diagnosis capture complementary properties. Detailed protocols
and case analyses are provided in the supplementary material.

\section{Conclusion}

We present AcoustiTrace, a diagnostic benchmark for evaluating acoustic
physical realism in T2AV and I2AV generation. Featuring eight dimensions
organized around the acoustic process and grounded in validated and
interpretable acoustic measurements, AcoustiTrace reveals that plausible
and synchronized sound events do not guarantee faithful modeling of the
tested acoustic relations. A targeted intervention further
demonstrates that diagnostic evidence from AcoustiTrace can guide model
refinement, providing a basis for incorporating acoustic principles into
training objectives, reward modeling, and candidate selection.

\clearpage
\setcounter{secnumdepth}{2}
\setcounter{section}{0}
\setcounter{subsection}{0}
\setcounter{subsubsection}{0}
\setcounter{figure}{0}
\setcounter{table}{0}
\setcounter{equation}{0}
\renewcommand{\thesection}{S\arabic{section}}
\renewcommand{\thesubsection}{\thesection.\arabic{subsection}}
\renewcommand{\thesubsubsection}{\thesubsection.\arabic{subsubsection}}
\renewcommand{\thefigure}{S\arabic{figure}}
\renewcommand{\thetable}{S\arabic{table}}
\renewcommand{\theequation}{S\arabic{equation}}

\begin{center}
  {\large\bfseries Supplementary Material for\\[2pt]
  AcoustiTrace: When Plausible Sound Violates Physics\par}
  \vspace{4pt}
  {\footnotesize Shiyang Li, Yuewen Cao, Yihao Liu, Yuandong Pu,\\
  Baochang Zhang, Xiaofei Li, Changqing Zou\par}
\end{center}
\vspace{6pt}

\begin{abstract}
This supplementary document clarifies the diagnostic abstraction and
positioning of AcoustiTrace and provides additional details on dataset
construction, evaluator operating domains, benchmark validity, evaluator
validation, comparison with PhyAVBench CPRS, and range-guided audio
sampling.
\end{abstract}

\section{Diagnostic Abstraction and Positioning}
\label{supp:diagnostic_abstraction}

AcoustiTrace treats each evaluation dimension as a relation-level
diagnostic contract rather than as an interchangeable output-quality
score. For dimension \(k\), we write this contract as
\begin{equation}
  \mathcal{D}_k
  =
  \bigl(
    \mathcal{V}_k,\mathcal{A}_k,\mathcal{R}_k,
    \mathcal{G}_k,\mathcal{S}_k
  \bigr),
  \label{eq:diagnostic_contract}
\end{equation}
where \(\mathcal{V}_k\) extracts the required visual evidence,
\(\mathcal{A}_k\) measures the corresponding audio quantity,
\(\mathcal{R}_k\) specifies the expected acoustic relation,
\(\mathcal{G}_k\) determines whether that relation is measurable in the
generated output, and \(\mathcal{S}_k\) maps the relation residual to an
interpretable diagnostic score. Learned detectors, trackers, depth
estimators, and MLLMs implement parts of the evidence-extraction
pipeline; the diagnostic target is defined by the acoustic relation,
validity conditions, and residual that connect the two modalities.

A central feature of this formulation is that the expected relation is
conditioned on evidence in the generated video whenever applicable,
rather than assuming that the model faithfully rendered the requested
event, trajectory, or environment. If the evidence required by
\(\mathcal{G}_k\) is absent or cannot be measured reliably, the output is
marked invalid for that dimension. If it is present, the evaluator
quantifies the deviation between the visual implication and the audio
measurement. This separates observability or prompt-compliance failures
from violations of an acoustic relation.

The shared contract gives the benchmark four complementary properties.
\textbf{Generated-output conditioning} ties evaluation to what the model
actually rendered in both modalities. \textbf{Mechanism localization}
places each diagnosed relation within sound generation, propagation
environment, or acoustic reception. \textbf{Validity-aware measurement}
reports coverage separately from conditional physical-consistency
scores. \textbf{Measurement-to-intervention} exposes relation residuals
that can serve as concrete optimization targets, as demonstrated by the
range-guided sampling study. Table~\ref{tab:diagnostic_contracts}
summarizes how all eight dimensions instantiate this abstraction.

\begin{table*}[t]
  \centering
  \begin{tabular}{@{}
    >{\raggedright\arraybackslash}p{0.13\textwidth}
    >{\raggedright\arraybackslash}p{0.15\textwidth}
    >{\raggedright\arraybackslash}p{0.19\textwidth}
    >{\raggedright\arraybackslash}p{0.17\textwidth}
    >{\raggedright\arraybackslash}p{0.25\textwidth}@{}}
    \toprule
    Acoustic stage & Dimension & Visual evidence & Audio quantity &
      Diagnosed relation \\
    \midrule
    Generation
      & Motion--Loudness
      & Localized action strength
      & Matched event level
      & Stronger visible excitation should emit greater acoustic energy. \\
    Generation
      & Log Attack Time
      & Object, material, and contact evidence
      & Attack duration
      & Onset dynamics should agree with the visible contact condition. \\
    Generation
      & Impact Decay
      & Isolated visible impact
      & Post-impact decay envelope
      & Impact energy should exhibit a physically plausible loss pattern. \\
    Propagation
      & RT60 Consistency
      & Scene geometry and material cues
      & Apparent 500 Hz RT60
      & Audio reverberation should agree with the visible environment. \\
    Propagation
      & Causality Violation
      & Visible event time
      & Audio-event onset
      & An audible effect should not precede its visible cause. \\
    Reception
      & Range Attenuation
      & Source-range trajectory
      & Short-time audio level
      & Level change should follow distance-dependent attenuation. \\
    Reception
      & Approach Gain
      & Decreasing source range
      & Short-time audio level
      & Approach toward the listener should produce a corresponding gain. \\
    Reception
      & Lateral Stability
      & Approximately constant range
      & Level drift and fluctuation
      & Level should remain stable without a substantial range change. \\
    \bottomrule
  \end{tabular}
  \caption{The common diagnostic contract instantiated by the eight
  AcoustiTrace dimensions. Each dimension links evidence in the generated
  video to a measurable audio quantity through an explicit acoustic
  relation and a dimension-specific validity gate.}
  \label{tab:diagnostic_contracts}
\end{table*}

This diagnostic unit complements existing evaluation paradigms.
General audio--video benchmarks such as VABench cover broad perceptual,
semantic, temporal, and content criteria \citep{hua2026vabench}.
Table~\ref{tab:benchmark_positioning} instead focuses on the
physics-oriented benchmarks most closely related to AcoustiTrace and
compares their task coverage and primary diagnostic capabilities. The
entries describe the protocols reported by the respective papers rather
than a ranking of their overall scope or quality.

The three task-coverage columns indicate whether a benchmark's primary
protocol directly evaluates text-to-audio--video (T2AV),
image-to-audio--video (I2AV), or video-to-audio (V2A) systems.
``Generated-video reference'' means that the expected physical relation
is inferred from visual evidence in the generated video, rather than
specified only by the prompt, an input video, or a real-world reference.
``Acoustic quantities'' indicates that evaluation is explicitly grounded
in measurable descriptors such as level, onset, decay, or reverberation,
rather than embedding similarity or generic rubric judgments alone.
``Per-output diagnosis'' requires a relation-specific score or violation
signal for an individual generated output, rather than only a pairwise
direction or a dataset-level statistic. ``Validity handling''
requires an explicit applicability gate that separates outputs lacking
the evidence needed for measurement from valid outputs with low
physical-consistency scores. ``Process-stage structure'' indicates that
the evaluated relations are organized across multiple stages of the
acoustic process rather than presented as an undifferentiated list.
Finally, ``metric-guided refinement'' requires the diagnostic quantity or
residual reported during evaluation to be reused as a generation or
optimization objective; incorporating physical conditioning without
closing this diagnostic loop is marked as partial support.

\begin{table*}[t]
  \centering
  \small
  \setlength{\tabcolsep}{2pt}
  \renewcommand{\arraystretch}{1.15}
  \begin{tabular}{@{}lccccccccc@{}}
    \toprule
    & \multicolumn{3}{c}{Task coverage} &
      \multicolumn{5}{c}{Diagnostic capabilities} &
      \multicolumn{1}{c}{Refinement} \\
    \cmidrule(lr){2-4}\cmidrule(lr){5-9}\cmidrule(l){10-10}
    Benchmark & T2AV & I2AV & V2A &
      \shortstack{Generated-video\\reference} &
      \shortstack{Acoustic\\quantities} &
      \shortstack{Per-output\\diagnosis} &
      \shortstack{Validity\\handling} &
      \shortstack{Process-stage\\structure} &
      \shortstack{Metric-guided\\refinement} \\
    \midrule
    PhyAVBench \citep{xie2025phyavbench}
      & \(\checkmark\) & \(\checkmark\) & \(\checkmark\)
      & \(\times\) & \(\times\) & \(\times\) & \(\times\)
      & \(\checkmark\) & \(\times\) \\
    FlatSounds \citep{li2026flatsounds}
      & \(\times\) & \(\times\) & \(\checkmark\)
      & \(\times\) & \(\checkmark\) & \(\triangle\) & \(\times\)
      & \(\triangle\) & \(\times\) \\
    AV-Phys Bench \citep{cui2026avphys}
      & \(\checkmark\) & \(\times\) & \(\times\)
      & \(\checkmark\) & \(\checkmark\) & \(\times\) & \(\times\)
      & \(\triangle\) & \(\times\) \\
    PAVAS / VGG-Impact \citep{oh2026pavas}
      & \(\times\) & \(\times\) & \(\checkmark\)
      & \(\times\) & \(\checkmark\) & \(\times\) & \(\times\)
      & \(\times\) & \(\triangle\) \\
    \midrule
    \textbf{AcoustiTrace}
      & \(\checkmark\) & \(\checkmark\) & \(\times\)
      & \(\checkmark\) & \(\checkmark\) & \(\checkmark\)
      & \(\checkmark\) & \(\checkmark\) & \(\checkmark\) \\
    \bottomrule
  \end{tabular}
  \caption{Capability matrix for representative physics-oriented
  audio--video benchmarks. \(\checkmark\) denotes direct support in the
  primary protocol, \(\triangle\) a related but partial capability or a
  different aggregation level, and \(\times\) that the capability is not
  part of the primary protocol.}
  \label{tab:benchmark_positioning}
\end{table*}

AcoustiTrace does not replace these complementary paradigms. Its
specific emphasis is the combination of generated-output conditioning,
mechanism-level organization, validity-aware scoring, and explicit
per-output relation diagnosis. These properties allow the same diagnostic residual
used for evaluation to be instantiated as a generation objective in the
range-guided study.

The remainder of this supplement documents the evidence supporting this
diagnostic contract. Real-world anchors establish relation recovery,
controlled perturbations test directional sensitivity, human judgments
assess preference alignment, the CPRS comparison tests metric
non-redundancy, and the range-guided study illustrates how a diagnosed
residual can be converted into an executable generation objective.

\section{Dataset Construction Details}
\label{supp:dataset}

AcoustiTrace combines real-world audio--video anchors with an
acoustically annotated RGB-D collection. The former support prompt
construction and evaluator validation under observable acoustic
relations, while the latter provide spatially resolved absorption cues
and RT60 supervision for the visual reverberation estimator.

\subsection{Real-World Audio--Video Anchors}
\label{supp:real_world_anchors}

We retrieve candidate videos from public sources using queries tailored
to the eight evaluation dimensions. The queries target different visible
motion strengths, isolated impacts, timed contacts, approaching or
receding sources, and motion at approximately constant range. For
Motion--Loudness, each case targets two visually distinguishable action
segments with different strengths. Using two actions limits
prompt-realization failures and ambiguity in event correspondence.
Impact Decay and Causality Violation use event-level clips centered on a
single visible contact or collision. Range Attenuation, Approach Gain,
and Lateral Stability use longer clips with a trackable source and a
clear increasing, decreasing, or approximately constant-range
trajectory.

Retrieval uses two task-oriented workflows. Receiver/observer queries
cover cars, motorcycles, trains, and airplanes with approaching,
receding, pass-by, and fly-over modifiers. Source-mechanics and causality
queries cover object drops, ball bounces, hammer and material impacts,
knocking, clapping, drum hits, breaking events, and everyday clicks or
closures. Curated seed categories additionally cover vehicle and
aircraft pass-bys, emergency sirens, drones or remote-controlled
vehicles, woodworking and metalworking, keyboard and switch clicks,
door mechanisms, sports impacts, and human percussive events.

Candidate clips undergo automatic screening followed by human review.
Media are standardized to 480p video and 16 kHz mono audio and processed
in 30 s chunks. Metadata filters remove sound-effect libraries, music,
remixes, compilations, and other unsuitable content; signal gates require
at least 3 dB RMS dynamic range and no more than 80\% silence. A
Qwen-family MLLM then checks source visibility, audio--visual
consistency, relative motion, event structure, camera motion, and
common contamination. Depending on the target dimension, the remaining
automatic stage combines open-vocabulary visual event localization,
audio event detection, source tracking, depth/applicability checks, and
signal-quality measurements. Source-event candidates use OV-AVEL
\citep{zhou2025ovavel} and FlexSED \citep{hai2025flexsed}, with visual
and audio events associated within 0.35 s.

Each retained clip is reviewed by two reviewers and kept only when both
agree that the target relation is observable in the video and that the
corresponding audio event or segment can be localized and measured.
Clips with ambiguous contacts, unstable trajectories, competing sound
sources, intrusive background music, excessive noise, or other
attribution confounds are excluded. The frozen collection contains
11,296 unique audio--video clips.

\paragraph{Role of the anchors and release scope}
The real-world anchors intentionally support both prompt construction
and evaluator validation. This alignment ensures that the evaluators'
declared operating domains cover the acoustic relations instantiated by
the prompt suites. Relation-recovery and controlled-perturbation
experiments therefore test measurability, sensitivity, and monotonicity
within that operating domain; they are not claims of open-domain
generalization, and the anchors are not generated samples used to rank
the evaluated models. For the public-video component, the release
manifest is limited to source video identifiers or URLs, time spans,
task labels, and derived annotations or scores; raw public-video media
are not redistributed.

\subsection{Acoustically Annotated RGB-D Observations}
\label{supp:annotated_rgbd}

The visual RT60 estimator is structured by the Sabine relation
\citep{kinsler2000fundamentals},
\begin{equation}
  T_{60}=0.161\frac{V}{A},
  \qquad
  A=\sum_j S_j\alpha_j,
  \label{eq:sabine}
\end{equation}
where \(V\) is room volume, \(S_j\) is the area of boundary region
\(j\), and \(\alpha_j\) is its frequency-dependent absorption
coefficient. Equation~\ref{eq:sabine} motivates a representation that
combines geometric information with spatially resolved material
absorption.

We first obtain RGB-D observations from Matterport3D
\citep{chang2017matterport3d}. GroundSAM segments visible surfaces and
objects, and a Qwen-family MLLM refines each coarse category into a
material description suitable for acoustic database retrieval. A
SentenceTransformer representation matches that description to its
top-1 PTB Room Acoustics Absorption Coefficient Database entry by cosine
similarity \citep{ptb_absorption_database}. We then project the matched
500 Hz absorption coefficient onto the corresponding mask; a missing
500 Hz coefficient in a matched PTB row is assigned 0.02. Combining all
regions produces a pixel-level Acoustic Alpha Map, which serves as a
third visual modality alongside RGB and depth. All final observations
use this semantic-material matching path; no HSV proxy or fallback maps
are included.

The 82,828 observations are split by sample-level stratification over
500 Hz RT60 bins into 57,977 training, 8,285 validation, and 16,566 test
samples. The split is not scene-disjoint: all three subsets contain
observations from the same 83 Matterport3D scenes. It is used for
in-domain estimator development and operating-domain coverage, rather
than as evidence of generalization to unseen scenes.
The frozen estimator-ablation artifact contains predictions for 16,563
of the 16,566 reconciled test-manifest observations; all ablation metrics
and visualizations below use this fixed scored subset.

For each visual observation, SoundSpaces~2.0
\citep{chen2022soundspaces2} simulates a room impulse response using
source and listener positions in the same Matterport3D scene. We filter
the impulse response in the octave band centered at 500 Hz, compute its
energy decay curve by Schroeder integration
\citep{schroeder1965reverberation}, and extrapolate the fitted decay
slope to obtain the reference RT60 label. The extraction converts the
RIR to mono at its original sampling rate, applies a fifth-order
Butterworth bandpass over 350--700 Hz, fits the Schroeder decay between
\(-5\) and \(-35\) dB, and computes \(T_{60}=-60/\mathrm{slope}\).
Across the final collection, the 500 Hz labels have mean 0.519 s,
standard deviation 0.342 s, median 0.375 s, and range 0.082--2.230 s.
The target is thus derived from the simulated acoustic response
independently of the visual inputs. These are simulated, derived labels
rather than in-situ measured room-acoustic ground truth. This procedure
produces 82,828 acoustically annotated RGB-D observations.

\begin{figure*}[t]
  \centering
  \includegraphics[width=\textwidth]{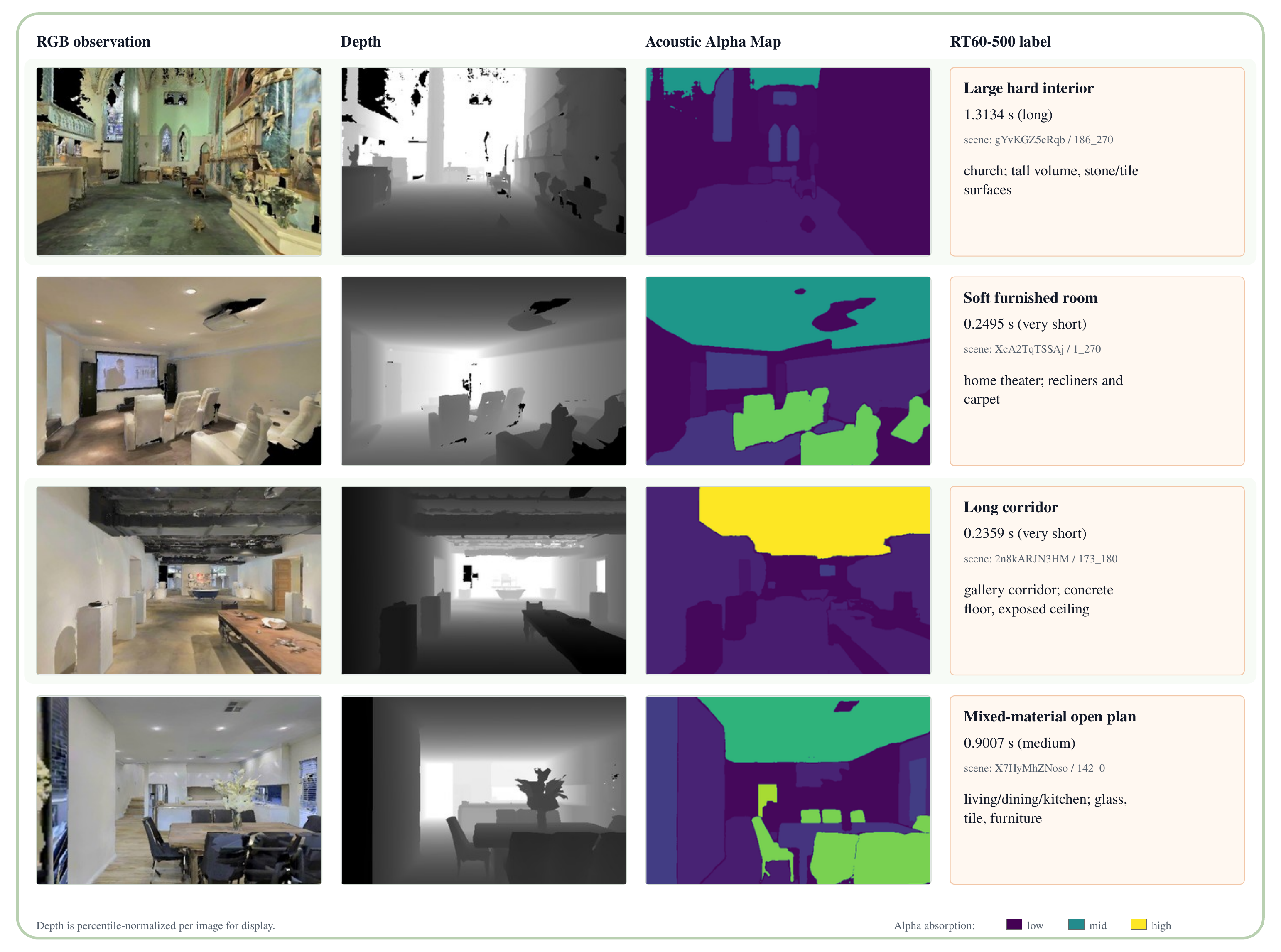}
  \caption{Representative acoustically annotated RGB-D observations.
  Each row shows an RGB observation, its aligned depth map, the derived
  500 Hz Acoustic Alpha Map, and the corresponding simulated 500 Hz RT60
  label. The examples illustrate variation in visible room geometry and
  spatially resolved surface absorption.}
  \label{fig:rt60_rgbd_gallery}
\end{figure*}

Figure~\ref{fig:rt60_rgbd_gallery} shows representative observations and
their aligned visual-acoustic annotations.

\section{Evaluator Definitions and Validity Criteria}
\label{supp:evaluators}

AcoustiTrace first determines whether the evidence required by an
evaluator is measurable and only then computes its physical-consistency
score. Invalid outputs are excluded from the corresponding conditional
mean rather than assigned a score of zero. We distinguish three kinds of
conditions throughout this section: programmatic checks implemented by
the evaluator, constraints encouraged by prompt or data construction,
and physical assumptions that delimit interpretation but are not
directly verified by code.
All reported evaluator scores are mapped to \([0,100]\), with larger
values indicating greater consistency for that dimension. These
normalized scores are dimension-specific summaries rather than a shared
physical unit and should not be compared as commensurate acoustic
quantities across dimensions.
Nevertheless, each evaluator is anchored by relation-consistent
real-world samples in the high-score regime and validated through
controlled perturbations that produce monotonic score degradation.
Cross-dimension score patterns can therefore be interpreted
descriptively as relative empirical difficulty within the calibrated
operating domains, but not as comparisons of absolute physical-error
magnitude.
The eight dimensions form a scoped set of measurable relations across
the three-stage acoustic process rather than an exhaustive inventory of
physical-acoustic phenomena. Future advances in detection and acoustic
measurement can extend both their operating domains and the relations
covered by the benchmark.

\subsection{Sound Generation}
\label{supp:sound_generation}

\subsubsection{Motion--Loudness}

Motion--Loudness requires localized visible actions and corresponding
audio events within the same generated sample. The visual branch of
OV-AVEL \citep{zhou2025ovavel} localizes candidate action intervals,
while FlexSED \citep{hai2025flexsed} detects candidate sound events.
Temporal association produces paired visual and audio segments. T2AV and
I2AV use the same visual-ranking path: a Qwen-based motion judgment with
a frame-difference fallback ranks the localized actions using visual
evidence only, considering motion amplitude, speed, range, and duration.
For each paired audio event, we compute the maximum short-time RMS level
in decibels within the matched window. The visual ordering and signed
audio-level difference reveal whether the stronger visible action is
paired with the louder event. A sample is invalid when the required
actions or their audio events cannot be localized or associated
reliably.
The event association permits at most \(0.35\) s audio--visual offset
and enforces at least \(0.12\) s between candidate events. Audio
segments extend from \(0.05\) s before to \(0.70\) s after the matched
event; the frame-difference fallback samples video at 8 fps in a
\([-0.25,0.10]\) s window around the visual peak. For each valid pair,
the no-margin indicator is one when the maximum RMS level of the
visually stronger action exceeds that of the weaker action. The
reported score is 100 times the mean of this indicator over valid
pairs.

\subsubsection{Log Attack Time}

Log Attack Time is evaluated only for I2AV. The conditioning image and
reference audio come from the same temporally aligned impact recording,
providing the object and material context for the generated sample. The
evaluator localizes the generated and reference impacts and computes a
smoothed RMS envelope for each matched window. A pre-onset interval
estimates the local baseline, and the transient peak defines the upper
end of the envelope range. Attack duration is the interval between the
first crossings of 10\% and 90\% of this baseline-to-peak range; taking
its logarithm yields Log Attack Time. The evaluator measures the
absolute difference between the generated and reference values. Audio
is analyzed at 22,050 Hz using 1,024-sample RMS frames, a 128-sample
hop, and Gaussian smoothing with \(\sigma=1\). The attack search window
extends from \(0.05\) s before to \(0.30\) s after onset, with the
preceding \(0.05\) s used to estimate the baseline and attack times
floored at \(10^{-4}\) s. For generated and reference attack durations
\(T_g\) and \(T_r\), the normalized score is
\begin{equation}
  S_{\mathrm{LAT}}
  =
  100\exp\left(
    -\frac{|\log T_g-\log T_r|}{0.35}
  \right).
  \label{eq:lat_score}
\end{equation}
The reference audio supplies the physical measurement and is not provided
to the generator. A sample is invalid if either impact cannot be
localized or its onset cannot be measured.

\subsubsection{Impact Decay}

Impact Decay applies to an isolated visible impact with a measurable
post-impact audio segment. The evaluator extracts an audio window around
the impact, computes a smoothed RMS envelope, locates the acoustic peak,
and treats the post-peak portion as the decay interval. Audio is
analyzed at 22,050 Hz in a segment from \(0.05\) s before to \(0.70\) s
after the event. The exponential decay fit uses the interval from
\(0.02\) to \(0.50\) s after the acoustic peak and includes a residual
energy floor. The segment must contain at least 64 samples, at least
five envelope points, and at least five post-peak fit points. Let
\(R^2\) be the decay-fit coefficient of determination,
\(E_{\mathrm{tail}}\) the tail residual mean absolute error in
decibels, and \(D_{\mathrm{pf}}\) the peak-to-floor dynamic range in
decibels. The per-event decay-shape score is
\begin{equation}
  S_{\mathrm{decay}}
  =
  100\,\operatorname{clip}(R^2,0,1)
  \exp\left(
    -\frac{E_{\mathrm{tail}}}
    {\max(D_{\mathrm{pf}},10^{-6})}
  \right).
  \label{eq:impact_decay_score}
\end{equation}
A steadily decreasing tail with a small residual receives a high score,
whereas abrupt truncation, sustained plateaus, secondary peaks, and
irregular fluctuations reduce it. Validity requires a matched impact,
sufficient usable decay support, and a finite decay-shape score. No
minimum \(R^2\) threshold is imposed: \(R^2\) contributes continuously
to the score rather than serving as a separate hard validity gate.
Overlapping events, missing impacts, or segments without a usable decay
interval are outside the operating domain.

\subsection{Propagation Environment}
\label{supp:propagation_environment}

\subsubsection{RT60 Consistency}

\paragraph{Design rationale}
The visual estimator does not treat RT60 as an unconstrained image
regression target. A Sabine-guided physics head captures the principal
relation among room geometry, surface area, and absorption, while a
regularized residual branch accounts for factors not visible from a
single view. The residual branch is not intended to recover complete
room-acoustic ground truth from unobserved scene content.

\paragraph{Visual-side supervision}
The training target combines Matterport3D geometry, the PTB-derived
Acoustic Alpha Map, and the 500 Hz RT60 label obtained from a
SoundSpaces~2.0 RIR. The visual estimator uses Qwen3-VL-8B-Instruct as
its backbone. Each sample contains an RGB image, a depth map, an
Acoustic Alpha Map, and a short task prompt. The three images are
processed using the native Qwen3-VL image representation. We take the
last-layer hidden states, apply attention-mask-aware mean pooling, and
obtain a 4096-dimensional scene representation \(h\).

\paragraph{Sabine-guided physics head}
The pooled representation is passed through LayerNorm and dropout
(\(p=0.1\)), followed by separate geometry, absorption, residual,
six-bin classification, and extreme-long heads. The geometry and
absorption branches predict positive latent quantities through
softplus:
\begin{align}
  \widehat V &= \operatorname{softplus}(f_V(h))+\epsilon, \\
  \widehat A_{500} &= \operatorname{softplus}(f_A(h))+\epsilon, \\
  \widehat T_{\mathrm{phys}}
    &= \operatorname{clip}\left(
      0.161\frac{\widehat V}{\widehat A_{500}},0.05,5.0
    \right).
\end{align}
The residual branch predicts a correction in log-RT60 space, giving
\begin{equation}
  \widehat T_{60}^{\mathrm{visual}}
  =
  \operatorname{clip}\left(
    \exp\left(\log\widehat T_{\mathrm{phys}}+f_{\mathrm{res}}(h)\right),
    0.05,5.0
  \right).
  \label{eq:visual_rt60_head}
\end{equation}
The model therefore returns \(\widehat V\), \(\widehat A_{500}\), the
Sabine prior, the residual, the final 500 Hz RT60 estimate, six RT60-bin
logits, and an extreme-long logit. The six bins are separated at
\(0.3\), \(0.6\), \(1.0\), \(1.5\), and \(2.0\) s.

\paragraph{Optimization}
The Qwen3-VL base weights are frozen. We train LoRA adapters and the
physics head; LoRA is applied to the query, key, value, output, gate,
up-projection, and down-projection modules with rank 16, scale 32, zero
dropout, and no bias parameters. The objective combines log-domain
SmoothL1 regression, six-bin cross entropy, extreme-long focal binary
cross entropy, physics consistency, residual \(L_1\) regularization,
and within-minibatch pairwise ranking:
\begin{equation}
  \begin{aligned}
    \mathcal L ={}&
    \mathcal L_{\mathrm{reg}}
    + 0.5\mathcal L_{\mathrm{class}}
    + \mathcal L_{\mathrm{long}} \\
    &+ 0.05\mathcal L_{\mathrm{phys}}
    + 0.01\mathcal L_{\mathrm{res}}
    + 0.1\mathcal L_{\mathrm{rank}}.
  \end{aligned}
\end{equation}
The regression tail weights for targets below 0.5 s, in
\([0.5,1.0)\), \([1.0,1.5)\), \([1.5,2.0)\), and at least 2.0 s are
1, 2, 4, 8, and 16, respectively. The class weights use the same
increasing-tail principle, with weights \([1,1,2,4,8,16]\). The
extreme-long focal term uses positive weight 16 and focusing parameter
2; the ranking margin is 0.1 with at most 512 pairs per minibatch.

Training uses three epochs, per-device batch size 2, gradient
accumulation over 16 steps, learning rate \(2\times10^{-4}\), weight
decay 0.01, warmup ratio 0.03, cosine scheduling, bf16, gradient
checkpointing, and maximum gradient norm 1. The image processor uses
12,544 and 401,408 as its minimum and maximum pixel budgets. Both the
direct SFT baselines and the physics-head model select their best
checkpoint using validation-set performance before evaluation on the
held-out simulated test split.

\paragraph{Direct SFT baselines}
The direct baselines use the same Qwen3-VL backbone and RGB, depth, and
Acoustic Alpha Map inputs, but autoregressively produce a JSON field
named \texttt{rt60\_500Hz} rather than using Equation~
\ref{eq:visual_rt60_head}. Their base weights are likewise frozen and
the same LoRA modules are trained. Standard direct SFT uses token-level
causal language-model cross entropy. Numeric-weighted SFT assigns
weight 5 to numeric target tokens and weight 1 to other supervised
tokens. The JSON baselines impose no model-side hard clamp; an
unparseable numeric field is treated as a parse failure.

\begin{table}[!t]
  \centering
  \begin{tabular*}{\columnwidth}{@{\extracolsep{\fill}}lrrrr@{}}
    \toprule
    Estimator & MAE & RMSE & \(\rho\) & Bias \\
    \midrule
    Vanilla & 1.2527 & 1.3532 & 0.2583 & 1.2488 \\
    Direct SFT & 0.0610 & 0.1009 & 0.8635 & -0.0223 \\
    Numeric SFT & \textbf{0.0601} & 0.0998 & 0.8696 & -0.0185 \\
    Sabine head & 0.0642 & \textbf{0.0930} &
      \textbf{0.8698} & \textbf{-0.0010} \\
    \bottomrule
  \end{tabular*}
  \caption{Visual RT60 estimation on the held-out simulated test split.
  Rows denote vanilla Qwen3-VL, direct JSON LoRA SFT, numeric-weighted
  LoRA SFT, and the Sabine-guided physics head. MAE, RMSE, and bias are
  in seconds; \(\rho\) is Spearman correlation. Checkpoints are selected
  using the validation split.}
  \label{tab:visual_rt60_ablation}
\end{table}

\begin{table}[!t]
  \centering
  \begin{tabular}{lr}
    \toprule
    Reference RT60 interval & Sabine-head MAE (s) \\
    \midrule
    \(0.0\)--\(0.3\) s & 0.0473 \\
    \(0.3\)--\(0.6\) s & 0.0551 \\
    \(0.6\)--\(1.0\) s & 0.0930 \\
    \(1.0\)--\(1.5\) s & 0.0988 \\
    \(>1.5\) s & 0.1106 \\
    \bottomrule
  \end{tabular}
  \caption{RT60-stratified error of the final visual estimator on the
  held-out simulated test split.}
  \label{tab:visual_rt60_ranges}
\end{table}

Table~\ref{tab:visual_rt60_ablation} shows that direct SFT reduces MAE
from 1.2527 to 0.0610, and numeric-weighted SFT attains the lowest MAE
of 0.0601. The Sabine-guided parameterization has an MAE of 0.0642 while
achieving the lowest RMSE (0.0930), near-zero bias (\(-0.0010\)), and a
Spearman correlation of 0.8698. Its error is higher in the shortest
RT60 interval but lower than numeric-weighted SFT above 1.5 s (0.1106
versus 0.1361 s), consistent with the physics head's emphasis on the
long-reverberation tail. The scored test subset contains only nine
examples with RT60 at least 2.0 s. The extreme-long head predicts six
such examples, with recall \(1/9\) and precision \(1/6\); these values
are reported as a tail diagnostic rather than evidence of robust
extreme-range performance.

\paragraph{Operating range and limitations}
The learned output is numerically bounded to \([0.05,5.0]\) s, whereas
the simulated labels used for training and evaluation span
approximately \(0.08\)--\(2.23\) s and are concentrated below 1.5 s.
The quantitative evidence therefore supports aggregate performance
over the simulated label range, with limited evidence in the
extreme-long tail. A single RGB-D observation cannot fully determine
out-of-view geometry, room openings, ceiling height, occluded
furnishings, or hidden materials. Moreover, the Acoustic Alpha Map is a
semantic-material absorption proxy rather than an in-situ measurement.
Real recordings and generated audio additionally introduce microphone,
source, rendering, post-processing, and dynamic-occlusion effects that
are absent from the simulated labels. The visual estimator is used
within this defined operating domain and is not presented as a
measurement-grade estimator for arbitrary real rooms.

The benchmark validity decision requires a present, parseable visual
prediction in the supported numeric range; it is not an explicit
open-set or out-of-distribution detector.

\paragraph{Audio-side apparent RT60}
The audio estimator identifies a usable decay interval and computes a
noise-compensated Schroeder energy decay curve after filtering the
matched waveform in the octave band centered at 500 Hz. It returns an
apparent RT60 only when the decay tail can be measured and fit. Repeated
event sequences, no valid event candidate, insufficient post-event tail,
or insufficient dynamic range can therefore invalidate the audio proxy.
A matched waveform is resampled to 24 kHz and filtered over
\([500/\sqrt{2},500\sqrt{2}]\) Hz. Candidate impulses require at least
10 dB peak prominence; the response starts 5 ms before the peak, and
the noise floor is estimated from 0.50 to 0.05 s before it. Noise-cutoff
margins of 10 dB, with a 6 dB fallback, are evaluated using 10 ms frames
and a 2 ms hop. A usable decay must span at least 0.050 s, each candidate
fit at least 0.020 s, and the candidate fit must attain
\(R^2\geq0.65\). T20, T10, and EDT candidates are considered in that
order, and finite audio proxies outside \([0.05,5.0]\) s are rejected.
A missing visual prediction or an incompatible audio--visual room class
can also invalidate the final comparison. Apparent RT60 estimated from
generated audio is conceptually distinct from room RT60 measured from a
controlled RIR. When both estimates are available, their discrepancy is
measured by the absolute log ratio
\(\left|\log(\widehat{T}_{60}^{\mathrm{audio}}/
\widehat{T}_{60}^{\mathrm{visual}})\right|\). Visual predictions outside
\([0.08,3.0]\) s are rejected. Writing
\(r=\widehat{T}_{60}^{\mathrm{audio}}/
\widehat{T}_{60}^{\mathrm{visual}}\), the consistency score is
\begin{equation}
  S_{\mathrm{RT60}}
  =
  100\,\operatorname{clip}\left(
    \frac{\log 3-|\log r|}{\log 3-\log 1.5},0,1
  \right).
  \label{eq:rt60_score}
\end{equation}
Thus ratios within a factor of 1.5 receive the maximum score and ratios
at or beyond a factor of 3 receive zero.

\subsubsection{Causality Violation}

Causality Violation requires a localized visible contact and a matched
audio event. The detected audio onset is compared with the visible
contact time. Under the canonical scoring convention, an onset earlier
than the visible event by more than 1 ms is counted as a causality
violation. Audio and visual detections are clustered with gaps of
0.20 s and 0.30 s, respectively, after filtering detections below
confidence 0.20. Normal-mode matching allows up to 0.08 s early,
0.25 s synchronous, and 0.70 s delayed evidence; these tolerances define
event association and are distinct from the 1 ms scoring threshold.
The 1 ms margin is the minimum timestamp increment used to implement a
strict temporal-precedence test; it does not imply 1 ms event-localization
accuracy.
For finite delays \(\delta_i=t_i^{\mathrm{audio}}-
t_i^{\mathrm{visual}}\) in valid matched events, the conditional score
is
\begin{equation}
  S_{\mathrm{causal}}
  =
  100\left(
    1-\frac{1}{n}\sum_{i=1}^{n}
    \mathbb{1}[\delta_i<-0.001]
  \right).
  \label{eq:causality_score}
\end{equation}
Samples without a reliable matched event pair are excluded from this
conditional score but remain in the denominator of the separately
reported output-level validity rate.

\subsection{Acoustic Reception}
\label{supp:acoustic_reception}

\subsubsection{Range Attenuation}

Range Attenuation evaluates relative range--level consistency within a
screened operating domain. In an ideal free field, the expected relative
level change is
\begin{equation}
  \Delta L^\star(t)
  =
  -20\log_{10}\frac{d(t)}{d(t_0)},
  \label{eq:range_reference}
\end{equation}
where \(d(t)\) is the relative source range. The visual branch tracks the
dominant visible source with Grounded-SAM, exports VDA-derived depth in
point-cloud form, and aggregates source-mask-supported points over time
to obtain a relative range trajectory. This trajectory is aligned with
the short-time audio level curve. Because source power can vary over a full clip, the
comparison is performed in local sliding windows. Within each window,
the observed relative level is compared with
Equation~\ref{eq:range_reference}, and \(R^2\) measures how well the
reference explains the generated level trajectory. Programmatic
screening checks track availability, depth validity, and applicability
of the range change. The implementation uses 0.40 s windows with a
0.05 s stride. Visual depth is median-smoothed over 0.35 s; a track
requires at least six finite depth points, at least 50\% finite support,
and a reliable run of at least 0.50 s, with gaps longer than 0.35 s
splitting a run. Range scoring requires at least four finite trajectory
points overall and at least three per scoring window. A dominant
receding segment must last at least 1.50 s and span at least 2.00 in the
evaluator's internal depth coordinate. These implementation thresholds
do not imply metric distances. The evaluator does not recover metric
source--receiver distance or assume that arbitrary generated scenes
satisfy ideal free-field conditions. Approximately stable source power,
source directionality, and environmental reflections remain modeling
assumptions rather than directly verified validity conditions.

\subsubsection{Approach Gain}

Approach Gain uses intervals in which the tracked source approaches the
listener and the relative visual range decreases sufficiently to be
measured. It evaluates the monotonic agreement between the source-range
trajectory and the short-time audio-level trajectory, together with the
orientation of the fitted loudness trend. Track, depth, and
applicability screening precede scoring; flattened, reversed, or
temporally inconsistent gain patterns receive lower scores. It uses the
same 0.40 s windows and 0.05 s stride as Range Attenuation. A candidate
window must span at least 0.15 in the evaluator's relative-depth
coordinate and have fitted depth slope below \(-0.002\); the dominant
approaching segment must last at least 1.50 s.

\subsubsection{Lateral Stability}

Lateral Stability applies when the tracked source remains at
approximately constant range or moves primarily across the listener's
field of view. After selecting intervals with limited relative-depth
variation, it measures systematic level drift and short-time loudness
variation. The score should be interpreted within this screened domain,
not as evidence that source power or the acoustic environment is
globally stationary. The same sliding-window configuration is used, and
candidate windows require relative-depth coefficient of variation at
most 0.12. Visual applicability additionally requires image-plane
bounding-box center displacement of at least 0.03 and path length of at
least 0.05.

\section{Prompt Suite Construction}
\label{supp:prompts}

The prompt suites contain 605 unique T2AV prompts and 748 unique I2AV
prompts. Pilot prompt development used LTX-2.3 and, for its supported
T2AV setting, JavisDiT++. The remaining model families entered the
evaluation only after the prompt suites and validity criteria were
frozen. Pilot refinement was limited to improving the observability and
measurability of the target acoustic relation rather than selecting
prompts according to a generator's physical score. The same final
prompts and criteria were applied to all supported models, including the
pilot models. The final reported outputs for LTX-2.3 and JavisDiT++ were
regenerated after the freeze.

Model-independent validity criteria mean that the required visual and
audio evidence, applicability thresholds, extraction-success gates, and
exclusion rules do not depend on model identity. Different models can
therefore have different valid rates, but no model receives customized
validity thresholds or prompt-specific exceptions.

The evaluator assignment counts are 84 for Range Attenuation, 222 each
for Approach Gain and Lateral Stability, 110 for Motion--Loudness, 111
for Impact Decay, 138 for Causality Violation, and 84 for RT60
Consistency in T2AV. I2AV uses the same assignment counts and adds 143
Log Attack Time assignments. These are overlapping evaluator
assignments rather than a partition of the unique prompt suites:
Approach Gain and Lateral Stability share a receiver-motion pool, while
Causality Violation reuses eligible prompts from the Motion--Loudness
and Impact Decay pools. Six prompts are also shared between the Range
Attenuation pool and the common Approach Gain/Lateral Stability pool.
Accordingly, the unique T2AV count is
\(84+222+110+111+84-6=605\); adding the 143 I2AV-only Log Attack Time
prompts gives 748 unique I2AV prompts.

Prompt construction targets the operating domain of each evaluator: for
example, isolated impacts for decay analysis, visible near--far motion
for Range Attenuation, and approximately constant-range motion for
Lateral Stability. Prompts specify the visual configuration needed to
measure a relation. For Motion--Loudness, the two-action design balances
physical comparability against the limited temporal
controllability of current generators; requiring additional actions at
specified times would make prompt realization and event correspondence
less reliable.

\begin{table*}[t]
  \centering
  \small
  \setlength{\tabcolsep}{4pt}
  \renewcommand{\arraystretch}{1.10}
  \begin{tabular}{@{}p{0.14\textwidth}p{0.54\textwidth}p{0.22\textwidth}@{}}
    \toprule
    Dimension & Representative prompt excerpt &
    Evidence controlled by the prompt \\
    \midrule
    Range Attenuation &
    A fixed street-side view with one car moving steadily from the
    foreground into the distance against a stable background. &
    Single trackable source and sustained receding trajectory. \\
    Approach Gain &
    A fixed roadside view with one race car moving from a distant
    position toward the camera. &
    Single trackable source and sustained approaching trajectory. \\
    Lateral Stability &
    A fixed exterior view with one source crossing the scene at
    approximately constant apparent range and size. &
    Image-plane motion with limited relative-depth variation. \\
    Motion--Loudness &
    A fixed workshop view in which the same person uses the same hammer
    on the same wooden object, first lightly and then more forcefully. &
    Two separated contacts differing primarily in visible strength. \\
    Impact Decay &
    A fixed workshop view with one visible hammer--chisel impact on a
    wooden board followed by a still interval and no second strike. &
    Isolated hard transient and measurable post-impact tail. \\
    Causality Violation &
    A fixed court view with clearly visible, temporally separated
    racket--ball contacts. &
    Localizable visual contact times and attributable sound events. \\
    Log Attack Time &
    An I2AV first frame showing a close-up metal locker and a paired
    reference recording containing one visible stick impact. &
    Material/contact context and a measurable reference attack. \\
    RT60 Consistency &
    A static wide indoor-room view with one short hand clap followed by
    several seconds of stillness. &
    Visible room geometry/materials and an isolated reverberant tail. \\
    \bottomrule
  \end{tabular}
  \caption*{Representative prompt excerpts for the eight dimensions.
  RT60 Consistency follows the final 84-prompt assignment.}
\end{table*}

\paragraph{Screening and quality control}
Candidate prompts are assembled through dimension-specific pipelines:
receiver-motion candidates for Range Attenuation, Approach Gain, and
Lateral Stability; source-action candidates for Motion--Loudness;
isolated-contact candidates for Impact Decay and Causality Violation;
indoor RGB-D views for RT60 Consistency; and paired first-frame/reference
records for Log Attack Time. Candidate rewriting retains only the visual
and event structure needed to make the target relation observable.
Receiver candidates are screened for an external fixed viewpoint, one
dominant trackable source, a stable background, and sufficient visible
range or image-plane motion. Interior or attached-camera views,
toy/miniature sources, competing sources, and trajectories without a
measurable range pattern are removed.

For source-action dimensions, prompts with missing contacts,
unattributable body motion, soft or quasi-static interactions, continuous
friction, or overlapping/repeated events are excluded. Impact Decay
additionally requires an isolated hard contact and an uninterrupted
post-impact interval. RT60 candidates require an indoor overview, a
static view, one short source event, and sufficient event-free tail;
views with cuts, camera motion, speech, music, or competing activity are
removed. I2AV Log Attack Time records must have a visible contact object,
a parseable paired reference recording, and a usable conditioning first
frame.

\paragraph{Deduplication and manifest checks}
Exact and normalized-text duplicates are removed within each candidate
pool before dimension-stratified subsampling. Semantically similar impact
candidates are additionally grouped by environment, subject, object,
interaction, and material so that a narrow interaction type cannot
dominate the retained pool. Intentionally retained seed or index
variants are handled during stratified sampling rather than counted as
independent semantic coverage. Final manifest checks verify required
fields, file availability, task/evaluator membership, and the assignment
overlaps described above. The manifest is then frozen before final
generation; no prompt is added, removed, or rewritten according to a
model's physical-consistency score.

\section{Benchmark Protocol and Additional Results}
\label{supp:benchmark}

\subsection{Model Configurations}
\label{supp:model_configurations}

We evaluate LTX-2.3, JavisDiT++, NAVA, Ovi, UniVerse-1, MOVA, Veo 3.1,
Seedance 2.0, and Wan 2.7 using each model's supported generation
settings. Log Attack Time is I2AV-only because the conditioning image
identifies the struck object and material context. The evaluated
JavisDiT++ release does not provide a native I2AV mode, while UniVerse-1
does not provide a native text-only T2AV mode; these task combinations
are therefore marked unavailable.

\begin{table*}[t]
  \centering
  \footnotesize
  \setlength{\tabcolsep}{4.2pt}
  \renewcommand{\arraystretch}{1.08}
  \begin{tabular}{lllcl}
    \toprule
    Model/configuration & Mode & Video output & Steps & Key generation settings \\
    \midrule
    LTX-2.3 distilled 1.1 & T2AV &
      \(512{\times}768\), 121 frames at 24 fps &
      Default & Distilled-pipeline defaults \\
    LTX-2.3 distilled 1.1 & I2AV &
      \(512{\times}768\), 121 frames at 24 fps &
      Default & First-frame conditioning, strength \(1.0\) \\
    JavisDiT-v1.0-jav & T2AV &
      240p, 9:16, 65 frames at 16 fps &
      50 & CFG \(=5.0\) \\
    NAVA FP8 & T2AV &
      \(1280{\times}704\), \(\sim6\) s at 24 fps &
      50 & 37-frame latent setting; fixed seed \(123\) \\
    NAVA FP8 & I2AV &
      \(1280{\times}704\), \(\sim6\) s at 24 fps &
      50 & 37-frame latent setting; video/audio guidance \(=3.0/2.0\) \\
    Ovi & T2AV, I2AV &
      \(720{\times}720\) &
      50 & UniPC, shift \(=5.0\), video/audio guidance \(=4.0/3.0\) \\
    UniVerse-1 & I2AV &
      \(256{\times}256\), 125 frames &
      35 & Video/audio guidance \(=1.2/5.0\) \\
    MOVA & T2AV, I2AV &
      \(352{\times}640\), 121 frames at 24 fps &
      50 & CFG \(=5.0\), sigma shift \(=5.0\) \\
    \bottomrule
  \end{tabular}
  \caption{Verified output and sampling settings for locally executed
  model--task configurations.}
  \label{tab:model_configurations}
\end{table*}

\subsection{Validity Rates and Exclusion Criteria}
\label{supp:validity_rates}

Tables~\ref{tab:t2av_coverage} and \ref{tab:i2av_coverage} report the
fraction of generated outputs that pass the applicability and
measurement checks for each evaluator. Scores in the main paper are
conditional means over these valid outputs.

\newcommand{\validcell}[3]{\shortstack{#1/#2\\(#3)}}

\begin{table*}[t]
  \centering
  \small
  \setlength{\tabcolsep}{4.0pt}
  \renewcommand{\arraystretch}{1.15}
  \begin{tabular}{lrrrrrrr}
    \toprule
    Model & Range & Approach & Lateral & Motion & Impact & Causality & RT60 \\
    \midrule
    LTX-2.3 &
      \validcell{84}{84}{100.00} &
      \validcell{222}{222}{100.00} &
      \validcell{222}{222}{100.00} &
      \validcell{107}{110}{97.27} &
      \validcell{108}{111}{97.30} &
      \validcell{138}{138}{100.00} &
      \validcell{57}{84}{67.86} \\
    JavisDiT++ &
      \validcell{82}{84}{97.62} &
      \validcell{222}{222}{100.00} &
      \validcell{222}{222}{100.00} &
      \validcell{82}{110}{74.55} &
      \validcell{56}{111}{50.45} &
      \validcell{138}{138}{100.00} &
      \validcell{35}{84}{41.67} \\
    NAVA &
      \validcell{84}{84}{100.00} &
      \validcell{222}{222}{100.00} &
      \validcell{222}{222}{100.00} &
      \validcell{105}{110}{95.45} &
      \validcell{106}{111}{95.50} &
      \validcell{138}{138}{100.00} &
      \validcell{38}{84}{45.24} \\
    Ovi &
      \validcell{65}{84}{77.38} &
      \validcell{220}{222}{99.10} &
      \validcell{222}{222}{100.00} &
      \validcell{103}{110}{93.64} &
      \validcell{107}{111}{96.40} &
      \validcell{93}{138}{67.39} &
      \validcell{34}{84}{40.48} \\
    UniVerse-1    & \multicolumn{7}{c}{Not available} \\
    MOVA &
      \validcell{59}{84}{70.24} &
      \validcell{222}{222}{100.00} &
      \validcell{222}{222}{100.00} &
      \validcell{68}{110}{61.82} &
      \validcell{97}{111}{87.39} &
      \validcell{70}{138}{50.72} &
      \validcell{41}{84}{48.81} \\
    Veo 3.1 &
      \validcell{84}{84}{100.00} &
      \validcell{222}{222}{100.00} &
      \validcell{222}{222}{100.00} &
      \validcell{110}{110}{100.00} &
      \validcell{111}{111}{100.00} &
      \validcell{138}{138}{100.00} &
      \validcell{49}{84}{58.33} \\
    Seedance 2.0 &
      \validcell{84}{84}{100.00} &
      \validcell{222}{222}{100.00} &
      \validcell{222}{222}{100.00} &
      \validcell{110}{110}{100.00} &
      \validcell{111}{111}{100.00} &
      \validcell{138}{138}{100.00} &
      \validcell{41}{84}{48.81} \\
    Wan 2.7 &
      \validcell{84}{84}{100.00} &
      \validcell{222}{222}{100.00} &
      \validcell{222}{222}{100.00} &
      \validcell{110}{110}{100.00} &
      \validcell{111}{111}{100.00} &
      \validcell{138}{138}{100.00} &
      \validcell{55}{84}{65.48} \\
    \bottomrule
  \end{tabular}
  \caption{T2AV validity counts and rates for each model--evaluator
  cell. Each cell reports \(n_{\mathrm{valid}}/n_{\mathrm{total}}\),
  with the percentage in parentheses. ``Not available'' indicates that
  the evaluated official release does not natively support the
  corresponding task.}
  \label{tab:t2av_coverage}
\end{table*}

\begin{table*}[t]
  \centering
  \small
  \setlength{\tabcolsep}{3.5pt}
  \renewcommand{\arraystretch}{1.15}
  \begin{tabular}{lrrrrrrrr}
    \toprule
    Model & Range & Approach & Lateral & Motion & Impact & Causality & LAT & RT60 \\
    \midrule
    LTX-2.3 &
      \validcell{83}{84}{98.81} &
      \validcell{220}{222}{99.10} &
      \validcell{222}{222}{100.00} &
      \validcell{94}{110}{85.45} &
      \validcell{109}{111}{98.20} &
      \validcell{138}{138}{100.00} &
      \validcell{143}{143}{100.00} &
      \validcell{57}{84}{67.86} \\
    JavisDiT++   & \multicolumn{8}{c}{Not available} \\
    NAVA &
      \validcell{83}{84}{98.81} &
      \validcell{221}{222}{99.55} &
      \validcell{221}{222}{99.55} &
      \validcell{83}{110}{75.45} &
      \validcell{109}{111}{98.20} &
      \validcell{138}{138}{100.00} &
      \validcell{143}{143}{100.00} &
      \validcell{47}{84}{55.95} \\
    Ovi &
      \validcell{60}{84}{71.43} &
      \validcell{219}{222}{98.65} &
      \validcell{220}{222}{99.10} &
      \validcell{82}{110}{74.55} &
      \validcell{105}{111}{94.59} &
      \validcell{66}{138}{47.83} &
      \validcell{143}{143}{100.00} &
      \validcell{53}{84}{63.10} \\
    UniVerse-1 &
      \validcell{65}{84}{77.38} &
      \validcell{214}{222}{96.40} &
      \validcell{219}{222}{98.65} &
      \validcell{56}{110}{50.91} &
      \validcell{53}{111}{47.75} &
      \validcell{87}{138}{63.04} &
      \validcell{143}{143}{100.00} &
      \validcell{40}{84}{47.62} \\
    MOVA &
      \validcell{43}{84}{51.19} &
      \validcell{102}{222}{45.95} &
      \validcell{150}{222}{67.57} &
      \validcell{89}{110}{80.91} &
      \validcell{89}{111}{80.18} &
      \validcell{87}{138}{63.04} &
      \validcell{143}{143}{100.00} &
      \validcell{40}{84}{47.62} \\
    Veo 3.1 &
      \validcell{84}{84}{100.00} &
      \validcell{222}{222}{100.00} &
      \validcell{222}{222}{100.00} &
      \validcell{110}{110}{100.00} &
      \validcell{111}{111}{100.00} &
      \validcell{138}{138}{100.00} &
      \validcell{143}{143}{100.00} &
      \validcell{45}{84}{53.57} \\
    Seedance 2.0 &
      \validcell{84}{84}{100.00} &
      \validcell{222}{222}{100.00} &
      \validcell{222}{222}{100.00} &
      \validcell{94}{110}{85.45} &
      \validcell{111}{111}{100.00} &
      \validcell{138}{138}{100.00} &
      \validcell{143}{143}{100.00} &
      \validcell{52}{84}{61.90} \\
    Wan 2.7 &
      \validcell{84}{84}{100.00} &
      \validcell{222}{222}{100.00} &
      \validcell{222}{222}{100.00} &
      \validcell{110}{110}{100.00} &
      \validcell{111}{111}{100.00} &
      \validcell{138}{138}{100.00} &
      \validcell{143}{143}{100.00} &
      \validcell{69}{84}{82.14} \\
    \bottomrule
  \end{tabular}
  \caption{I2AV validity counts and rates for each model--evaluator
  cell. Each cell reports \(n_{\mathrm{valid}}/n_{\mathrm{total}}\),
  with the percentage in parentheses. LAT denotes Log Attack Time.}
  \label{tab:i2av_coverage}
\end{table*}

\paragraph{Aggregate invalid coverage}
Across the 120 supported model--task--evaluator cells (56 T2AV and 64
I2AV), the benchmark contains 16,680 output--evaluator measurement
attempts. Of these, 649 of 7,768 T2AV attempts (8.35\%) and 993 of 8,912
I2AV attempts (11.14\%) are invalid.

\paragraph{Invalid-reason audit}
We audit the available per-output reason and status fields using a
conservative taxonomy. Evidence tags can overlap, and a record is left
as unknown when the available fields establish invalidity but do not
support a more specific attribution. The audit is used to explain
failure mechanisms; exact validity counts remain those in
Tables~\ref{tab:t2av_coverage} and \ref{tab:i2av_coverage}.

\textbf{Range Attenuation.}
Most invalid Range records lack a finite primary score without exposing
a more specific failure reason and are therefore retained as unknown. A
smaller set explicitly indicates missing audio or an unavailable visual
track/depth input.

\textbf{Approach Gain.}
Invalid outputs primarily fail the visual applicability gate because an
approaching interval, depth trajectory, or reliable track cannot be
verified. Missing or near-silent audio and extraction failures account
for additional I2AV cases.

\textbf{Lateral Stability.}
All T2AV outputs pass the Lateral Stability gate. The I2AV invalid cases
mainly reflect unverifiable constant-range/lateral applicability or
missing audio, with ambiguous records retained as unknown.

\textbf{Motion--Loudness.}
Both tasks use the same Qwen-based motion judgment and frame-difference
fallback. Invalidity arises when this path cannot establish a reliable
visual ordering, when too few actions or audio events can be localized
and matched, or when the generation or evaluator input is incomplete.

\textbf{Impact Decay.}
The main failure modes are a missing or unmatched impact, insufficient
post-impact support for a finite decay fit, and failed or incomplete
evaluator input. As described above, a successful finite fit is required
but \(R^2\) is not used as an additional hard threshold.

\textbf{Causality Violation.}
Outputs are invalid when the evaluator finds too few matched
audio--visual events or clusters, substantial unmatched event evidence,
or no finite delay estimate.

\textbf{RT60 Consistency.}
The dominant evidence concerns an unavailable audio RT60 proxy, including
repeated event sequences, no valid event candidate, insufficient
post-event tail, or insufficient dynamic range. Missing visual
predictions and audio--visual room-class mismatches can overlap with
these audio-side failures.

\subsection{Conditional Means and Bootstrap Confidence Intervals}
\label{supp:bootstrap_ci}

Each prompt yields exactly one generated output for a given model.
Within each model--task--evaluator cell, we retain the valid generated
outputs and draw 10,000 bootstrap resamples of the same size with
replacement. We recompute the conditional mean for every resample and
report the 2.5th and 97.5th percentiles as the 95\% confidence interval.
Resampling is performed independently within each cell using a fixed
cell-specific random seed. We do not apply prompt-level clustering
because there are no repeated outputs for the same prompt within a
model. Tables~\ref{tab:t2av_bootstrap_ci} and
\ref{tab:i2av_bootstrap_ci} report the resulting intervals.

\newcommand{\bootci}[3]{\shortstack{#1\\{[#2--#3]}}}

\begin{table*}[t]
  \centering
  \small
  \setlength{\tabcolsep}{2.2pt}
  \renewcommand{\arraystretch}{1.15}
  \begin{tabular}{@{}lccccccc@{}}
    \toprule
    Model & Range & Approach & Lateral & Motion & Impact & Causality & RT60 \\
    \midrule
    LTX-2.3 &
      \bootci{81.63}{77.78}{85.37} &
      \bootci{74.11}{71.82}{76.29} &
      \bootci{92.18}{91.42}{92.91} &
      \bootci{88.51}{82.24}{94.12} &
      \bootci{90.00}{86.87}{92.86} &
      \bootci{86.52}{80.57}{92.03} &
      \bootci{41.08}{34.18}{48.30} \\
    JavisDiT++ &
      \bootci{39.17}{33.96}{44.69} &
      \bootci{68.39}{65.75}{71.05} &
      \bootci{87.55}{86.46}{88.53} &
      \bootci{51.52}{40.54}{62.20} &
      \bootci{73.62}{64.73}{81.73} &
      \bootci{86.60}{80.45}{92.03} &
      \bootci{69.81}{60.84}{78.46} \\
    NAVA &
      \bootci{68.12}{63.30}{72.84} &
      \bootci{69.88}{67.65}{72.12} &
      \bootci{84.42}{82.85}{85.83} &
      \bootci{75.86}{67.29}{83.81} &
      \bootci{88.59}{85.78}{91.08} &
      \bootci{93.22}{88.87}{97.10} &
      \bootci{69.74}{62.02}{77.36} \\
    Ovi &
      \bootci{57.13}{51.25}{62.52} &
      \bootci{70.62}{68.18}{72.97} &
      \bootci{89.56}{88.43}{90.64} &
      \bootci{56.44}{46.73}{66.02} &
      \bootci{98.62}{94.59}{100.00} &
      \bootci{88.89}{82.44}{94.62} &
      \bootci{69.07}{60.37}{77.31} \\
    UniVerse-1 & \multicolumn{7}{c}{Not available} \\
    MOVA &
      \bootci{62.42}{56.09}{68.70} &
      \bootci{81.14}{76.99}{84.91} &
      \bootci{86.80}{84.99}{88.57} &
      \bootci{74.00}{63.24}{83.82} &
      \bootci{59.10}{52.61}{65.52} &
      \bootci{83.33}{74.28}{91.43} &
      \bootci{53.57}{46.42}{60.79} \\
    Veo 3.1$^{*}$ &
      \bootci{71.85}{65.87}{77.43} &
      \bootci{88.68}{81.74}{94.81} &
      \bootci{81.79}{72.55}{90.21} &
      \bootci{75.10}{66.56}{82.93} &
      \bootci{98.61}{97.96}{99.15} &
      \bootci{98.62}{96.52}{100.00} &
      \bootci{73.51}{64.92}{81.61} \\
    Seedance 2.0$^{*}$ &
      \bootci{86.48}{80.50}{92.52} &
      \bootci{71.32}{66.07}{75.52} &
      \bootci{76.57}{75.44}{77.65} &
      \bootci{83.98}{76.71}{90.34} &
      \bootci{91.39}{85.16}{97.02} &
      \bootci{99.16}{97.48}{100.00} &
      \bootci{24.28}{17.70}{31.26} \\
    Wan 2.7$^{*}$ &
      \bootci{74.49}{66.36}{82.61} &
      \bootci{71.69}{67.56}{75.00} &
      \bootci{82.03}{72.80}{89.05} &
      \bootci{82.37}{75.10}{89.09} &
      \bootci{85.91}{72.65}{97.66} &
      \bootci{95.68}{92.06}{98.58} &
      \bootci{33.20}{26.71}{39.83} \\
    \bottomrule
  \end{tabular}
  \caption{T2AV conditional means with percentile 95\% bootstrap
  confidence intervals over valid generated outputs. All values are on
  the 0--100 scale.}
  \label{tab:t2av_bootstrap_ci}
\end{table*}

\begin{table*}[t]
  \centering
  \small
  \setlength{\tabcolsep}{1.8pt}
  \renewcommand{\arraystretch}{1.15}
  \begin{tabular}{@{}lcccccccc@{}}
    \toprule
    Model & Range & Approach & Lateral & Motion & Impact & Causality & LAT & RT60 \\
    \midrule
    LTX-2.3 &
      \bootci{66.89}{62.08}{71.63} &
      \bootci{65.22}{62.76}{67.57} &
      \bootci{91.33}{90.04}{92.41} &
      \bootci{72.73}{63.82}{81.63} &
      \bootci{84.95}{81.12}{88.45} &
      \bootci{88.39}{82.61}{93.46} &
      \bootci{62.56}{58.01}{67.16} &
      \bootci{42.12}{34.67}{49.67} \\
    JavisDiT++ & \multicolumn{8}{c}{Not available} \\
    NAVA &
      \bootci{67.78}{63.07}{72.52} &
      \bootci{65.88}{63.08}{68.71} &
      \bootci{86.91}{85.48}{88.19} &
      \bootci{89.55}{82.71}{95.57} &
      \bootci{83.64}{79.85}{87.20} &
      \bootci{87.80}{82.00}{92.87} &
      \bootci{65.29}{60.33}{70.39} &
      \bootci{70.71}{64.08}{77.17} \\
    Ovi &
      \bootci{45.30}{39.55}{51.11} &
      \bootci{71.98}{69.57}{74.25} &
      \bootci{91.08}{90.03}{92.00} &
      \bootci{54.33}{43.90}{64.76} &
      \bootci{83.20}{80.23}{86.15} &
      \bootci{50.60}{38.48}{62.12} &
      \bootci{50.42}{44.06}{56.80} &
      \bootci{66.91}{60.11}{73.49} \\
    UniVerse-1 &
      \bootci{32.53}{20.81}{40.67} &
      \bootci{65.81}{63.47}{68.22} &
      \bootci{84.70}{83.38}{85.87} &
      \bootci{50.77}{37.50}{63.27} &
      \bootci{63.13}{53.93}{71.48} &
      \bootci{44.97}{34.48}{55.31} &
      \bootci{34.65}{29.63}{39.73} &
      \bootci{42.06}{34.87}{49.30} \\
    MOVA &
      \bootci{53.50}{46.03}{61.21} &
      \bootci{78.79}{73.84}{83.29} &
      \bootci{86.01}{83.32}{88.38} &
      \bootci{80.90}{73.01}{88.76} &
      \bootci{84.28}{80.30}{87.77} &
      \bootci{90.00}{83.45}{95.75} &
      \bootci{63.88}{59.83}{67.84} &
      \bootci{55.01}{45.75}{63.92} \\
    Veo 3.1$^{*}$ &
      \bootci{63.70}{48.69}{74.37} &
      \bootci{85.76}{81.46}{90.06} &
      \bootci{81.14}{77.13}{84.52} &
      \bootci{84.53}{77.27}{90.91} &
      \bootci{83.48}{77.04}{91.12} &
      \bootci{97.36}{94.46}{99.53} &
      \bootci{73.28}{68.32}{78.36} &
      \bootci{46.17}{38.07}{54.58} \\
    Seedance 2.0$^{*}$ &
      \bootci{59.98}{51.58}{69.01} &
      \bootci{70.83}{67.80}{73.86} &
      \bootci{79.50}{74.81}{82.76} &
      \bootci{83.92}{76.23}{91.00} &
      \bootci{87.39}{83.56}{90.89} &
      \bootci{97.03}{94.06}{99.28} &
      \bootci{50.15}{45.19}{55.23} &
      \bootci{17.22}{12.29}{22.64} \\
    Wan 2.7$^{*}$ &
      \bootci{45.17}{36.55}{53.70} &
      \bootci{76.12}{73.04}{80.09} &
      \bootci{77.38}{63.37}{90.28} &
      \bootci{85.67}{79.04}{91.82} &
      \bootci{72.46}{62.20}{87.21} &
      \bootci{95.75}{92.22}{98.65} &
      \bootci{73.70}{68.74}{78.78} &
      \bootci{76.28}{71.08}{81.44} \\
    \bottomrule
  \end{tabular}
  \caption{I2AV conditional means with percentile 95\% bootstrap
  confidence intervals over valid generated outputs. LAT denotes Log
  Attack Time. All values are on the 0--100 scale.}
  \label{tab:i2av_bootstrap_ci}
\end{table*}

\subsection{Matched-Valid Analysis}
\label{supp:matched_valid}

Conditional scoring can favor a low-coverage model if only its easiest
outputs remain valid. A matched-valid analysis addresses this concern by
retaining only prompts for which every compared model yields a valid
score. For each evaluator, we intersect the valid prompt sets of the
compared models and recompute their conditional means on that shared
subset. We analyze LTX-2.3, JavisDiT++, and Ovi for T2AV, and LTX-2.3,
Ovi, and MOVA for I2AV. These groups include a high-coverage reference
and models with lower coverage on different evaluators. The resulting
percentile 95\% confidence intervals use 10,000 prompt-level bootstrap
resamples.

\begin{table*}[t]
  \centering
  \small
  \setlength{\tabcolsep}{7.0pt}
  \renewcommand{\arraystretch}{1.15}
  \begin{tabular}{@{}lcccc@{}}
    \toprule
    Evaluator & \(n_{\mathrm{common}}\) & LTX-2.3 & JavisDiT++ & Ovi \\
    \midrule
    Range Attenuation & 62 &
      \bootci{80.96}{74.88}{86.60} &
      \bootci{40.08}{32.59}{47.94} &
      \bootci{57.74}{50.71}{64.50} \\
    Approach Gain & 220 &
      \bootci{74.64}{72.00}{77.24} &
      \bootci{68.86}{65.74}{71.90} &
      \bootci{70.18}{67.37}{72.87} \\
    Lateral Stability & 222 &
      \bootci{92.04}{91.15}{92.89} &
      \bootci{87.28}{86.06}{88.40} &
      \bootci{89.18}{87.85}{90.39} \\
    Motion--Loudness & 78 &
      \bootci{87.96}{80.12}{94.50} &
      \bootci{52.74}{42.28}{63.19} &
      \bootci{58.62}{50.78}{66.46} \\
    Impact Decay & 52 &
      \bootci{89.34}{83.26}{94.17} &
      \bootci{72.84}{65.19}{79.76} &
      \bootci{97.16}{90.41}{100.00} \\
    Causality Violation & 93 &
      \bootci{86.92}{80.07}{93.77} &
      \bootci{87.18}{80.39}{93.97} &
      \bootci{89.12}{82.79}{95.45} \\
    RT60 Consistency & 30 &
      \bootci{42.16}{29.31}{55.61} &
      \bootci{68.92}{55.23}{80.89} &
      \bootci{69.48}{51.79}{82.39} \\
    \bottomrule
  \end{tabular}
  \caption{T2AV matched-valid results on the intersection of prompts
  valid for LTX-2.3, JavisDiT++, and Ovi. Each model cell reports the
  conditional mean above and percentile 95\% bootstrap confidence
  interval below. All scores are on the 0--100 scale.}
  \label{tab:t2av_matched_valid}
\end{table*}

Within this three-model comparison, the ordering is unchanged from
Table~\ref{tab:t2av_bootstrap_ci} for Range Attenuation, Approach Gain,
Lateral Stability, Motion--Loudness, Impact Decay, and Causality
Violation. RT60 Consistency is the only exception: JavisDiT++ and Ovi
exchange their closely matched order, with strongly overlapping
confidence intervals, while LTX-2.3 remains substantially lower. The
principal cross-model patterns therefore persist after holding the valid
prompt set fixed, although these results should not be interpreted as a
single global leaderboard.

For I2AV, each bootstrap draw resamples the common prompt IDs and applies
the same sampled IDs to all three models. Table~\ref{tab:i2av_matched_valid}
reports the resulting matched-valid means and intervals.

\begin{table*}[t]
  \centering
  \small
  \setlength{\tabcolsep}{7.0pt}
  \renewcommand{\arraystretch}{1.15}
  \begin{tabular}{@{}lcccc@{}}
    \toprule
    Evaluator & \(n_{\mathrm{common}}\) & LTX-2.3 & Ovi & MOVA \\
    \midrule
    Range Attenuation & 43 &
      \bootci{66.73}{59.42}{73.76} &
      \bootci{45.82}{39.13}{52.61} &
      \bootci{53.50}{47.97}{58.38} \\
    Approach Gain & 102 &
      \bootci{65.41}{62.45}{68.44} &
      \bootci{72.16}{69.71}{74.61} &
      \bootci{78.79}{76.35}{81.26} \\
    Lateral Stability & 150 &
      \bootci{91.20}{90.21}{92.20} &
      \bootci{90.94}{89.92}{91.97} &
      \bootci{86.01}{84.90}{87.13} \\
    Motion--Loudness & 82 &
      \bootci{72.22}{62.55}{81.88} &
      \bootci{54.44}{43.55}{65.33} &
      \bootci{80.57}{72.07}{89.07} \\
    Impact Decay & 89 &
      \bootci{84.81}{81.68}{87.86} &
      \bootci{83.36}{79.80}{86.70} &
      \bootci{84.28}{81.27}{87.34} \\
    Causality Violation & 66 &
      \bootci{87.96}{80.44}{95.49} &
      \bootci{50.45}{38.44}{62.46} &
      \bootci{89.48}{81.97}{95.49} \\
    Log Attack Time & 143 &
      \bootci{62.56}{58.27}{66.82} &
      \bootci{50.42}{46.04}{54.90} &
      \bootci{63.88}{59.53}{68.15} \\
    RT60 Consistency & 40 &
      \bootci{42.45}{34.79}{50.07} &
      \bootci{66.54}{58.35}{74.88} &
      \bootci{55.01}{46.78}{63.26} \\
    \bottomrule
  \end{tabular}
  \caption{I2AV matched-valid results on the intersection of prompts
  valid for LTX-2.3, Ovi, and MOVA. Each model cell reports the
  conditional mean above and prompt-level bootstrap 95\% confidence
  interval below. All scores are on the 0--100 scale.}
  \label{tab:i2av_matched_valid}
\end{table*}

The matched-valid ordering is identical to the original conditional
ordering in Table~\ref{tab:i2av_bootstrap_ci} for all eight I2AV
evaluators. Across the 24 model--evaluator means, restricting evaluation
to the common prompt sets changes the reported mean by at most 0.52
points. Thus, within this targeted robustness comparison, the principal
I2AV rankings are not explained by differences in validity coverage.

\section{Evaluator Validation}
\label{supp:validation}

We validate the evaluators with three complementary forms of evidence.
Screened real-world anchors test whether the intended relation can be
recovered when it is present, controlled single-factor perturbations test
whether the score responds to a targeted violation, and paired human
judgments test whether the resulting preference agrees with expert
perception. No single external dataset provides numerical ground truth for
all eight relations, so each dimension is paired with the strongest
available reference appropriate to its native acoustic quantity.

\subsection{Real-World Relation Recovery}
\label{supp:relation_recovery}

We first apply each evaluator to unmodified real-world audio--video
anchors for which the target relation is observable and measurable.
These samples define the unperturbed operating point for the controlled
sweeps below. This validation checks relation recovery within the
screened domains; it does not claim absolute calibration for arbitrary
open-domain video.

Relation-recovery evidence takes two forms. When an independent numerical
comparator is available, as for paired Log Attack Time references or the
audio- and visual-side RT60 proxies described below, we compare the
recovered acoustic quantities directly. For the remaining dimensions,
the two-reviewer screening procedure confirms before scoring that the
anchor contains the required event and relation, such as two visibly
different action strengths, an isolated impact, a receding or approaching
source, or motion at approximately constant range. The purpose is to
verify that an evaluator recovers the expected relation when its required
evidence is observable, rather than to require every open-domain recording
to receive the maximum score.

\begin{figure*}[!t]
  \centering
  \includegraphics[width=\textwidth]
    {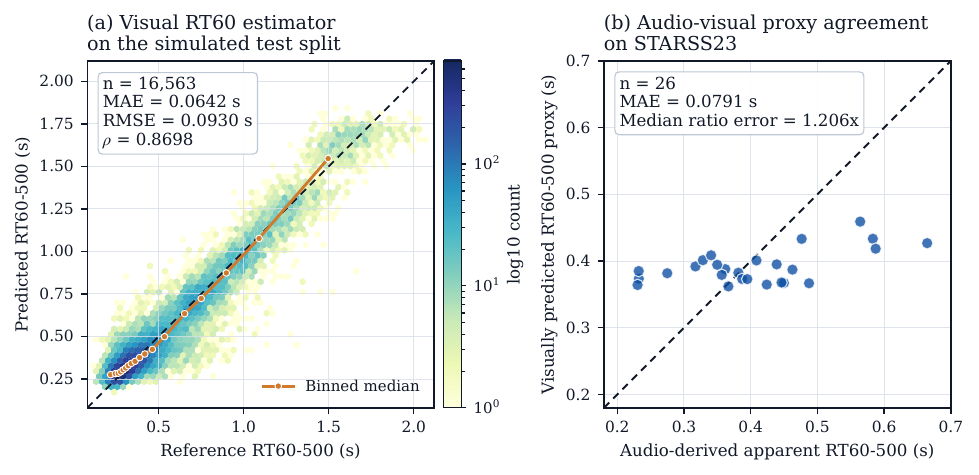}
  \caption{Visual RT60 validation. \textbf{(a)} Predictions from the
  Sabine-guided physics head on the 16,563 scored observations in the
  held-out simulated test subset. Hexagons show sample density, the
  orange curve shows binned medians, and the dashed line denotes
  identity. \textbf{(b)} Agreement between independently audio-derived
  apparent RT60 and visually predicted RT60 proxies for the 26 valid
  STARSS23 clapping windows. STARSS23 supplies event annotations rather
  than room-level RT60 ground truth; both axes in panel (b) are proxies.}
  \label{fig:rt60_validation}
\end{figure*}

\subsection{Controlled Perturbation Experiments}
\label{supp:perturbations}

Starting from the real-world anchors, we selectively perturb the audio
quantity targeted by each evaluator while leaving the corresponding
visual evidence unchanged. Motion--Loudness attenuates the event
associated with the stronger visible action by up to 12 dB. Log Attack
Time scales attack duration by up to \(4\times\). Impact Decay modifies
the decay-tail profile, while RT60 Consistency changes the apparent
audio-to-visual RT60 ratio away from one. Causality Violation advances
the audio onset by up to 120 ms. Range Attenuation and Approach Gain
progressively flatten the expected range-dependent envelope, and Lateral
Stability injects up to 8 dB of level fluctuation. Across all eight
dimensions, evaluator scores move in the expected direction as the
targeted violation becomes more severe.

The perturbations preserve the video stream, visible events, scene
semantics, and media duration. The audio transformation is constructed to
change the targeted relation while retaining the event needed by the
evaluator; the ordinary validity gates reject conditions in which that
event can no longer be measured comparably. For confidence intervals and
within-source dose-response summaries, the resampling unit is the source
anchor rather than an individual perturbed condition, so multiple
severity levels derived from the same recording are not treated as
independent samples.

The endpoint changes in the severity sweeps plotted in Figure 6(a) of the
main paper provide a direct summary for five dimensions. Relative to the
unmodified condition, 12 dB attenuation of the stronger-action event
reduces the mean Motion--Loudness score by 65.0 points. The strongest
decay-tail perturbation reduces Impact Decay by 33.0 points, complete
range-envelope flattening reduces Range Attenuation by 21.8 points and
Approach Gain by 13.1 points, and 8 dB injected fluctuation reduces
Lateral Stability by 34.1 points. The final clean-anchor Causality
Violation sweep retains 40 sources that are common-valid across the tested
onset advances; its mean score decreases as the audio is advanced from 0
to 120 ms.

\paragraph{Log Attack Time dose response}
The strict final analysis starts from 44 high-score identity anchors and
retains the 17 sources that remain valid at attack-duration factors
\(1\), \(1.5\), \(2\), and \(4\). On this common-valid subset, the median
within-source Spearman correlation between stretch factor and measured
Log Attack Time is 0.7746, with a source-bootstrap 95\% confidence
interval of [0.6325, 0.9487]. The corresponding correlation between
perturbation severity and consistency score is \(-0.7746\)
[-0.9487, -0.4000]. A \(4\times\) stretch reduces the score by 34.68
points on average [25.57, 43.33], with degradation in 82.4\% of retained
anchors (one-sided Wilcoxon \(p=6.1\times10^{-5}\)).

The response is deliberately interpreted at the evaluator's temporal
resolution rather than as a continuously strict curve. Its envelope hop
is 5.805 ms, and 35 of the 44 identity anchors occupy the same two-hop
attack-time bin. Consequently, 31 of the 51 adjacent common-valid
comparisons are quantization ties, while only 2 of 51 are true reversals.
This resolution audit explains the staircase response in Figure 6(a)
without treating unchanged adjacent bins as independent failures.

\paragraph{RT60 mismatch dose response}
For controlled RT60 validation, we fix the visual inputs and visual RT60
proxy for 26 anchors and replace only the audio with a fixed dry impact
convolved with synthetic 500 Hz room impulse responses. Six target
audio-to-visual RT60 ratios, \(0.50\), \(0.67\), \(1.00\), \(1.50\),
\(2.00\), and \(3.00\), produce 156 valid conditions with no skipped
estimates. In log-ratio space, the measured ratio tracks the injected
target with slope 0.9808, intercept 0.0544, \(R^2=0.8743\), and mean
absolute log residual 0.1750.

More importantly, the median within-anchor Spearman correlation between
injected mismatch severity and consistency score is \(-0.7827\), with a
source-bootstrap 95\% confidence interval of [-0.8874, -0.7392]. At
\(2\times\) and \(3\times\) mismatch, all 26 anchors decrease in score,
with median drops of 44.22 and 100.00 points, respectively. Ratios near
\(0.67\), \(1.00\), and \(1.50\) frequently saturate by design because
Equation~\ref{eq:rt60_score} assigns the maximum score within the
predefined \(1.5\times\) tolerance region.

Together, relation recovery and controlled perturbation test validity in
complementary directions: the former asks whether an expected acoustic
relation can be recovered when it is present and measurable, whereas the
latter asks whether a targeted violation is detected when the visual
content and non-target semantics are held fixed.

\begin{figure*}[!t]
  \centering
  \includegraphics[width=\textwidth]
    {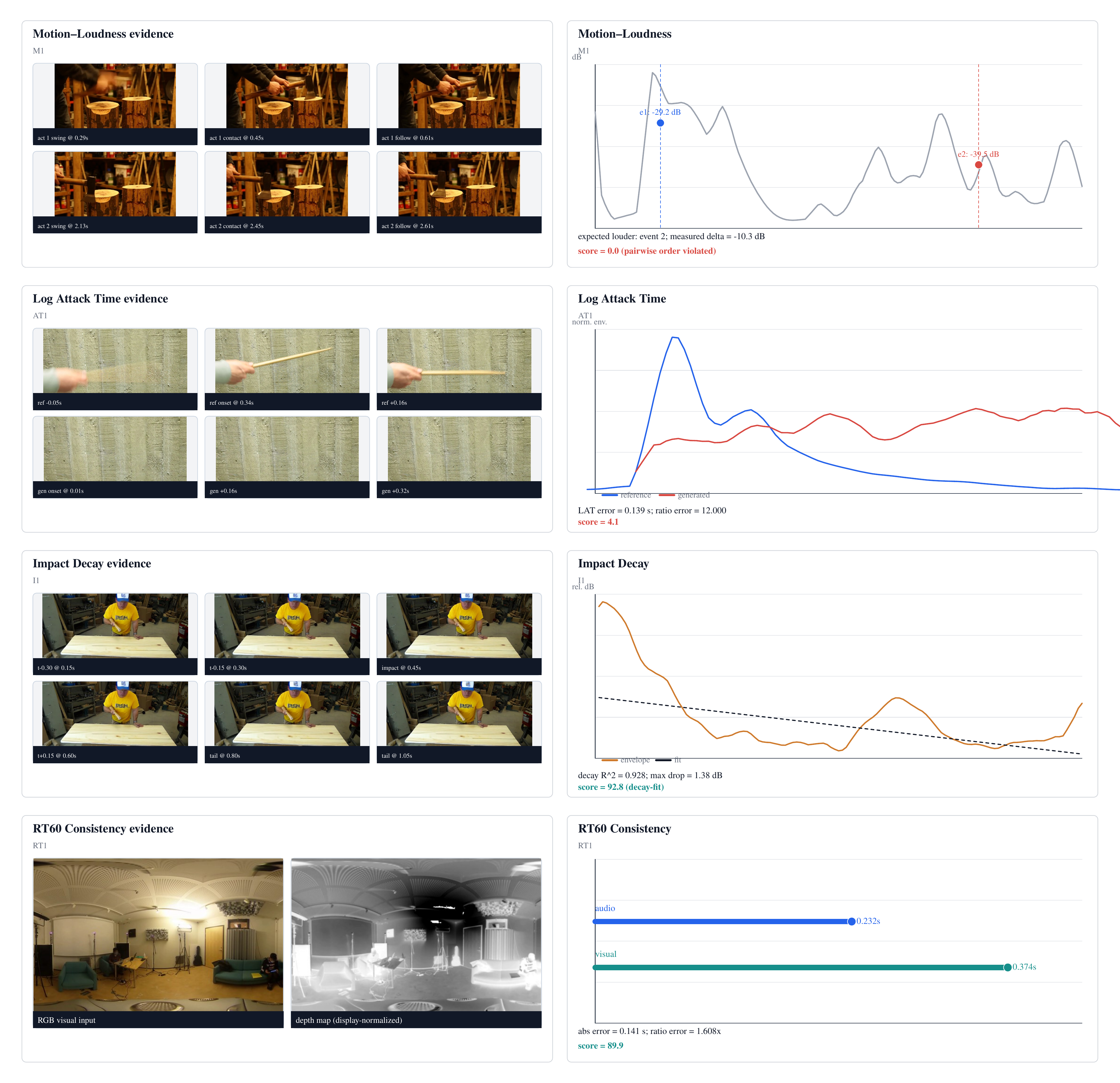}
  \caption{Representative diagnostic readouts for Motion--Loudness, Log
  Attack Time, Impact Decay, and RT60 Consistency. Each row pairs the
  evaluator's required evidence with the corresponding native acoustic
  measurement and final score.}
  \label{fig:evaluator_readouts_generation}
\end{figure*}

\subsection{Proxy Agreement on STARSS23}
\label{supp:starss23_validation}

We further evaluate RT60 Consistency on real indoor recordings from
STARSS23, the Sony-TAu Realistic Spatial Soundscapes 2023 dataset
\citep{starss23record,dcase2023starss23}. STARSS23 provides synchronized
multichannel audio and 360-degree video from real rooms, together with
temporal sound-event labels, directions of arrival, and source
distances. Its development data cover 16 rooms and 13 target event
classes, including Clapping. These annotations identify candidate
clapping intervals; they do not provide RT60 labels and are not treated
as room-acoustic ground truth in this experiment.

We extract 44 candidate clapping windows using the dataset-provided
event class and temporal annotations. We then apply the same 500 Hz
audio-side decay analysis used by the benchmark to derive an apparent
RT60 from the recorded waveform. Windows without an isolated,
measurable decay are rejected, primarily because of strong background
noise or insufficient energy in the 500 Hz band. This leaves 26 valid
windows. For each valid window, the visual RT60 estimator predicts a
500 Hz RT60 proxy from the temporally aligned video observation. The
analysis therefore compares an independently derived audio proxy with a
model-predicted visual proxy.

Writing \(T_a\) and \(T_v\) for the audio and visual proxies, we report
\begin{align}
  e_{\mathrm{abs}} &= |T_a-T_v|, \\
  e_{\log} &= |\log T_a-\log T_v|, \\
  e_{\mathrm{ratio}} &= \max(T_a/T_v,T_v/T_a).
  \label{eq:starss23_proxy_errors}
\end{align}
We also apply the benchmark consistency mapping in
Equation~\ref{eq:rt60_score}. Table~\ref{tab:starss23_rt60} summarizes
the resulting proxy agreement.

\begin{table}[t]
  \centering
  \begin{tabular}{lr}
    \toprule
    Statistic & Value \\
    \midrule
    Candidate clapping windows & 44 \\
    Valid paired proxies & 26 (59.1\%) \\
    Mean absolute error & 0.0791 s \\
    Median absolute error & 0.0739 s \\
    Mean absolute log error & 0.2013 \\
    Median ratio error & \(1.2063\times\) \\
    Mean consistency score & 98.56 \\
    Median consistency score & 100.00 \\
    Minimum consistency score & 85.51 \\
    \bottomrule
  \end{tabular}
  \caption{Agreement between audio-derived and visually predicted 500 Hz
  RT60 proxies on the valid STARSS23 clapping subset.}
  \label{tab:starss23_rt60}
\end{table}

Of the 26 valid pairs, 10 have absolute error at most 0.05 s, 17 at
most 0.10 s, and 23 at most 0.15 s. Thirteen pairs are within a factor
of \(1.20\times\), 18 are within \(1.25\times\), and all 26 are within
\(1.75\times\). The median ratio error of approximately
\(1.21\times\), together with a mean absolute error of 0.079 s,
summarizes agreement between the two proxies on this screened
real-recording subset.

\subsection{Measured-RIR Case Study on BRAS CR3}
\label{supp:bras_cr3}

We additionally examine one measured-room case from the Benchmark for
Room Acoustical Simulation (BRAS) \citep{brinkmann2021bras,aspoeck2020bras}.
BRAS was created to support reproducible room-acoustics comparisons and
provides documented room geometry, surfaces, source--receiver
configurations, and measured acoustic parameters. We select its CR3
complex music-room scene because it complements the event-centered
STARSS23 analysis: CR3 provides a measured RIR-derived reverberation
comparator, while its room volume, stage, wood floor, hard boundaries,
and seating supply diverse visual geometry and absorption cues.

CR3 was not used for RT60 training, hyperparameter tuning, or checkpoint
selection. The visual pipeline receives an RGB view, a
Video-Depth-Anything depth estimate, and a Grounded-SAM material
segmentation mapped to a 500 Hz Acoustic Alpha Map using the same
image-derived procedure as the benchmark; it does not use BRAS surface
descriptions or measured acoustics as input. The visual estimator predicts
1.1019 s in the 500 Hz octave band. The official BRAS file reports a
room-level 500 Hz T20 of 1.2584 s, without identifying that value as a mean
or median over individual RIRs. Their absolute difference is 0.1565 s
(12.4\% relative error; ratio error \(1.142\times\)). This case therefore
shows that the visual pipeline recovers a plausible measured-room
reverberation scale from independently inferred geometry and material cues.

BRAS categorizes CR1--CR4 as complex scenes with greater measurement
uncertainty and cautions against treating them as direct reference scenes.
Accordingly, we use CR3 as a single illustrative measured-RIR agreement
case, not as an aggregate external-validation estimate or evidence of
generalization across BRAS rooms.

\begin{figure}[!t]
  \centering
  \includegraphics[width=\columnwidth]
    {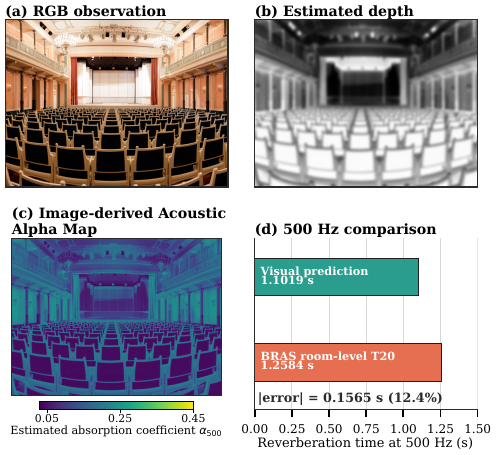}
  \caption{Measured-RIR case study on BRAS CR3. The visual estimator
  receives an RGB observation, independently estimated depth, and an
  image-derived 500 Hz Acoustic Alpha Map. CR3 was excluded from training,
  tuning, and checkpoint selection, and no BRAS surface descriptions or
  measured acoustics were used as model input. The prediction is 1.1019 s,
  compared with the official room-level 500 Hz T20 of 1.2584 s. This
  single-room result is illustrative rather than an aggregate BRAS
  evaluation.}
  \label{fig:bras_cr3_case}
\end{figure}

\begin{figure*}[!t]
  \centering
  \includegraphics[width=0.90\textwidth]
    {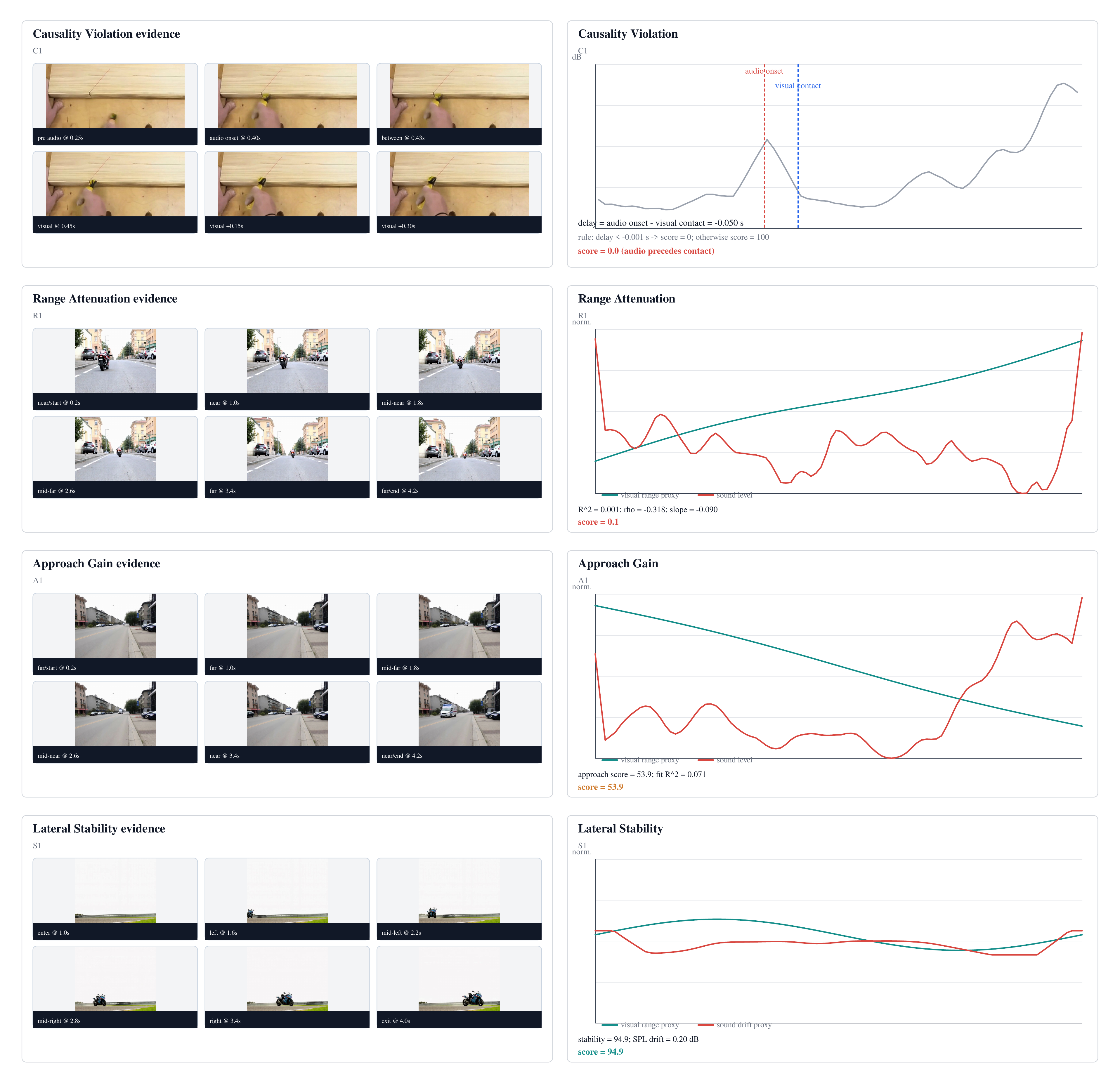}
  \caption{Representative diagnostic readouts for Causality Violation,
  Range Attenuation, Approach Gain, and Lateral Stability. Visual-event or
  range evidence is shown at left, and the corresponding temporal or
  level-trajectory diagnostic is shown at right.}
  \label{fig:evaluator_readouts_reception}
\end{figure*}

\subsection{Human Evaluation Protocol}
\label{supp:human_evaluation}

Controlled sensitivity alone does not establish that an evaluator's
preference matches human perception of the target relation. We therefore
test human alignment using cross-model outputs matched by prompt, task, and
dimension; real-world anchors and perturbed validation samples are
excluded. Each dimension has eight prompts and eight unique pairs, one
pair per prompt. Each of the 30 expert raters judges every pair exactly
once, yielding 30 ratings per pair and 1,920 judgments in total. Model
identities and evaluator scores are hidden,
left--right order is randomized, and raters judge only the target
acoustic relation using A, B, or Tie.
Table~\ref{tab:human_agreement} reports agreement between evaluator
preferences and human judgments.

\begin{table}[t]
  \centering
  \begin{tabular}{lc}
    \toprule
    Evaluation dimension & Agreement [95\% CI] (\%) \\
    \midrule
    Motion--Loudness  & 87.9 [81.41, 92.66] \\
    Log Attack Time   & 93.5 [88.01, 96.72] \\
    Impact Decay      & 89.5 [83.21, 93.86] \\
    RT60 Consistency  & 88.7 [82.31, 93.26] \\
    Causality Violation & 93.5 [88.01, 96.72] \\
    Range Attenuation & 91.1 [85.16, 95.07] \\
    Approach Gain     & 90.3 [84.26, 94.46] \\
    Lateral Stability & 92.7 [87.11, 96.27] \\
    \bottomrule
  \end{tabular}
  \caption{Agreement between evaluator preferences and human judgments
  on paired generated A/V samples, with 95\% confidence intervals.}
  \label{tab:human_agreement}
\end{table}

\subsection{Representative Evaluator Readouts}
\label{supp:diagnostic_readouts}

Figures~\ref{fig:evaluator_readouts_generation} and
\ref{fig:evaluator_readouts_reception} make the native diagnostic outputs
concrete with one screened example per dimension. In each row, the left
panel shows the visual or paired-reference evidence used by the evaluator,
and the right panel shows its event markers, fitted quantity, trajectory,
or consistency comparison. Scores are reported on the benchmark's 0--100
scale. These examples illustrate the evaluator readouts and are not
additional aggregate estimates.

\section{Comparison with PhyAVBench CPRS}
\label{supp:cprs}

PhyAVBench evaluates several physical relations in generated
audio--video and its official CPRS metric measures whether the embedding
transition between a paired condition follows a learned reference
direction \citep{xie2025phyavbench}. We use its inverse-square-law T34
subset because distance-dependent attenuation is the direct conceptual
overlap with AcoustiTrace Range Attenuation. Holding the official prompts
and near--far pairing fixed isolates the difference between an
embedding-direction metric and a diagnostic that compares the generated
visual distance change with the accompanying audio-level change.

\subsection{Official T34 Subset and Pair Generation}
\label{supp:cprs_protocol}

We use five fixed near--far prompt pairs from PhyAVBench's official T34
inverse-square-law subset \citep{xie2025phyavbench}. The subset is
selected from the official prompt file before generation rather than
post-selected from model outputs. Within each pair, the scene, source,
action, and fixed-camera description remain unchanged; only camera
distance changes. Side A is the near condition and side B is the far
condition. Table~\ref{tab:t34_prompt_subset} lists the complete subset.

For every model and prompt pair, we use seeds 0--19 and apply the same
seed to the near and far sides. The pairing key is
\((\text{model},\text{sample ID},\text{seed})\). This produces
\(5\times20=100\) near--far pairs and 200 generated videos per model.
LTX-2.3, NAVA, and JavisDiT++ therefore contribute 100 pairs each, for
300 pairs and 600 videos in total. Generation manifests verify that all
300 expected pairs contain both nonempty video files.

After generation, AcoustiTrace does not assume that a model obeyed the
camera distance written in the prompt. For each A/B video, VDA-derived
point-cloud export, Grounded-SAM source detection, and depth-track
aggregation estimate the visible source distance. These estimates define
\begin{equation}
  \Delta L_{\mathrm{expected}}
  =
  -20\log_{10}(d_B/d_A).
  \label{eq:t34_expected_delta}
\end{equation}
We extract mono audio at 16 kHz, compute global RMS levels, and obtain
\(\Delta L_{\mathrm{observed}}=L_B-L_A\). Their absolute difference and
the official near--far ordering provide the scalar AcoustiTrace score;
event windows for discrete sources and active windows for continuous
sources are retained as diagnostics. CPRS is computed independently on
the same fixed pairs using only the official LAION-CLAP backend. The
final table joins the two metrics by model, sample ID, seed, and A/B
video paths, with no ImageBind scores included.

\subsection{Model-Wise Correlation}
\label{supp:cprs_correlations}

Because the relation-matched overlap comprises five fixed official
prompt groups with 20 seeds per model, these correlations describe
repeated generations on this fixed T34 subset rather than a broader
prompt population. The reported Fisher intervals are pair-level
descriptive intervals conditional on this fixed subset.

Across all 300 pairs, CPRS and AcoustiTrace exhibit only weak negative
correlation. The model-wise estimates remain small and vary in sign or
uncertainty: LTX-2.3 is near zero, NAVA has a weak negative Pearson
correlation, and JavisDiT++ has a weak negative Spearman correlation.
The result is therefore not driven by one model and does not support
treating either metric as a substitute for the other.

\begin{figure*}[!t]
  \centering
  \includegraphics[width=\textwidth]{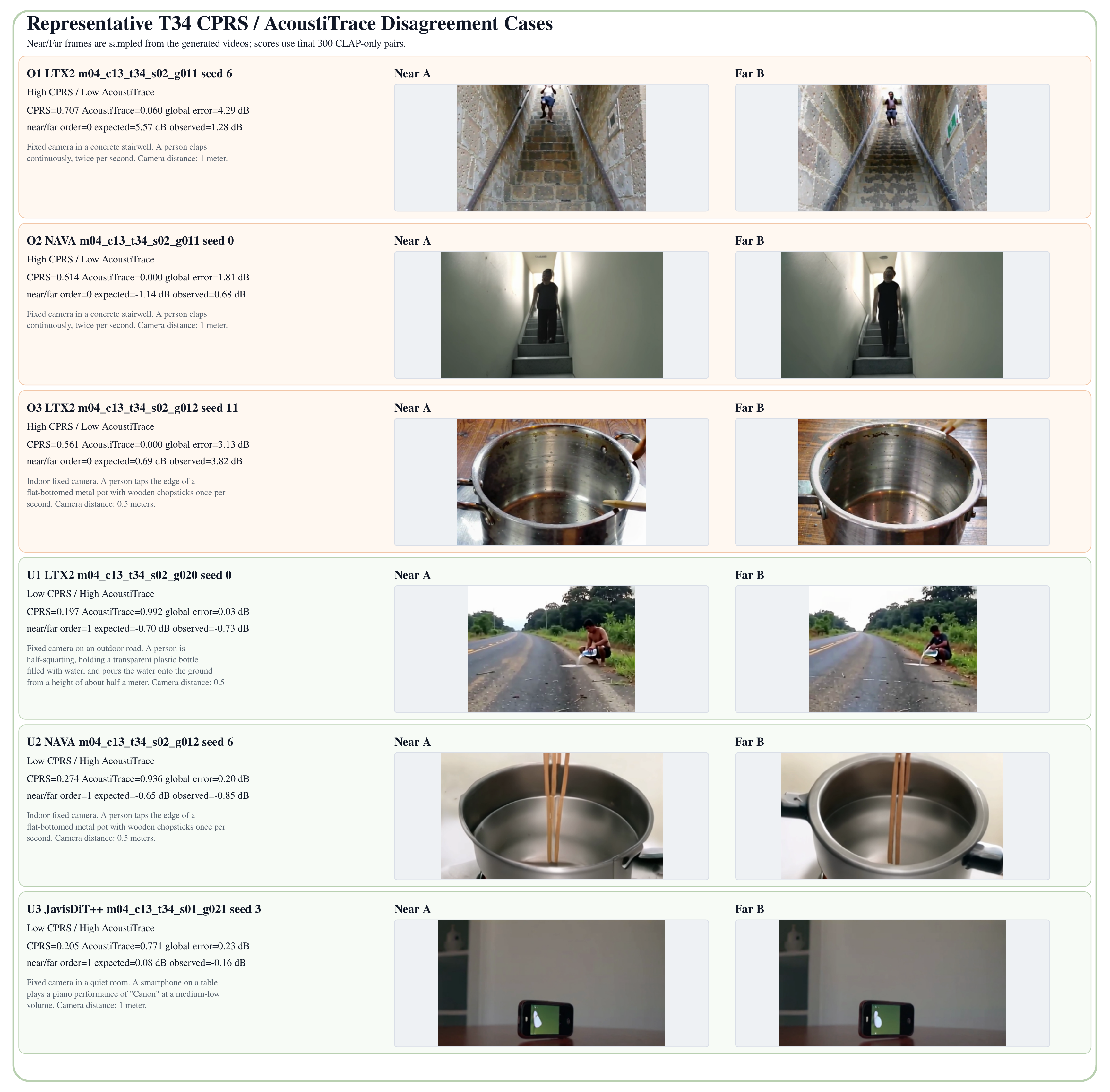}
  \caption{Representative CPRS--AcoustiTrace disagreement cases. Each
  row shows sampled near/far frames and the corresponding full-pair
  metrics. O1--O3 illustrate embedding over-credit; U1--U3 illustrate
  embedding under-credit relative to the explicit acoustic relation.}
  \label{fig:cprs_case_gallery}
\end{figure*}

\subsection{Representative Disagreement Cases}
\label{supp:cprs_cases}

We define high and low scores using 0.5 thresholds on both axes and
select three examples from each disagreement quadrant. Table~
\ref{tab:cprs_cases} reports the final selected cases. Here
\(\mathrm{Err}_{\Delta L}=
|\Delta L_{\mathrm{observed}}-\Delta L_{\mathrm{expected}}|\), and
Order equals one when the generated far-side audio is quieter than the
near-side audio under the official A/B assignment.

\begin{table*}[!t]
  \centering

  \begin{tabular*}{\textwidth}{@{\extracolsep{\fill}}llrrr@{}}
    \toprule
    Official sample ID & Source event & Near (m) & Far (m) &
      Prompt \(\Delta L\) (dB) \\
    \midrule
    \texttt{m04\_c13\_t34\_s02\_g012} & Metal-pot tapping & 0.5 & 5 & -20.00 \\
    \texttt{m04\_c13\_t34\_s02\_g011} & Clapping & 1 & 5 & -13.98 \\
    \texttt{m04\_c13\_t34\_s01\_g021} & Smartphone piano & 1 & 5 & -13.98 \\
    \texttt{m04\_c13\_t34\_s02\_g020} & Pouring water & 0.5 & 8 & -24.08 \\
    \texttt{m04\_c13\_t34\_s03\_g013} & Outdoor speech & 0.5 & 8 & -24.08 \\
    \bottomrule
  \end{tabular*}
  \caption{Official T34 near--far prompt subset. The prompt-level
  expected change is \(-20\log_{10}(d_{\mathrm{far}}/
  d_{\mathrm{near}})\) and describes the generation condition; final
  AcoustiTrace scoring instead uses distance estimated from each
  generated video pair.}
  \label{tab:t34_prompt_subset}

  \vspace{0.6\baselineskip}

  \begin{tabular*}{\textwidth}{@{\extracolsep{\fill}}lrll@{}}
    \toprule
    Model & \(n\) & Pearson \(r\) [95\% CI] &
      Spearman \(\rho\) [95\% CI] \\
    \midrule
    Overall & 300 & -0.134 [-0.243, -0.021] &
      -0.142 [-0.251, -0.029] \\
    LTX-2.3 & 100 & 0.005 [-0.192, 0.201] &
      0.075 [-0.123, 0.267] \\
    NAVA & 100 & -0.228 [-0.406, -0.033] &
      -0.188 [-0.371, 0.008] \\
    JavisDiT++ & 100 & -0.166 [-0.351, 0.031] &
      -0.264 [-0.437, -0.071] \\
    \bottomrule
  \end{tabular*}
  \caption{Correlation between LAION-CLAP CPRS and AcoustiTrace over the
  final fixed T34 pairs. Confidence intervals use the Fisher
  \(z\)-transform approximation.}
  \label{tab:cprs_model_correlations}

  \vspace{0.6\baselineskip}

  \begin{tabular*}{\textwidth}{@{\extracolsep{\fill}}lllrrrrr@{}}
    \toprule
    Case & Model & Official sample ID & Seed & CPRS & AcoustiTrace &
      \(\mathrm{Err}_{\Delta L}\) (dB) & Order \\
    \midrule
    O1 & LTX-2.3 & \texttt{m04\_c13\_t34\_s02\_g011} & 6 &
      0.707 & 0.060 & 4.29 & 0 \\
    O2 & NAVA & \texttt{m04\_c13\_t34\_s02\_g011} & 0 &
      0.614 & 0.000 & 1.81 & 0 \\
    O3 & LTX-2.3 & \texttt{m04\_c13\_t34\_s02\_g012} & 11 &
      0.561 & 0.000 & 3.13 & 0 \\
    U1 & LTX-2.3 & \texttt{m04\_c13\_t34\_s02\_g020} & 0 &
      0.197 & 0.992 & 0.03 & 1 \\
    U2 & NAVA & \texttt{m04\_c13\_t34\_s02\_g012} & 6 &
      0.274 & 0.936 & 0.20 & 1 \\
    U3 & JavisDiT++ & \texttt{m04\_c13\_t34\_s01\_g021} & 3 &
      0.205 & 0.771 & 0.23 & 1 \\
    \bottomrule
  \end{tabular*}
  \caption{Representative disagreements from the final 300 pairs.
  O1--O3 are high-CPRS/low-AcoustiTrace cases; U1--U3 are
  low-CPRS/high-AcoustiTrace cases.}
  \label{tab:cprs_cases}
\end{table*}

All three O cases violate the official near--far audio ordering and
have attenuation errors from 1.81 to 4.29 dB despite receiving CPRS
above 0.5. Conversely, all three U cases preserve the ordering and
match the visually inferred attenuation within 0.23 dB, yet receive
CPRS below 0.5. Figure~\ref{fig:cprs_case_gallery} shows sampled frames
from the paired videos; the metrics are computed from the complete
generated A/V pairs rather than from these frames. These complementary
errors explain the weak aggregate correlation: CPRS measures an
embedding transition along its official reference direction, whereas
AcoustiTrace tests the generated pair's explicit visual-distance and
audio-attenuation relation.

\section{Range-Guided Audio Sampling Details}
\label{supp:range_guidance}

The benchmark's main purpose is diagnosis, but a useful diagnostic should
also expose a residual that can guide generation. We choose Range
Attenuation for this proof of concept because its visual range trajectory
and audio envelope define a continuous, differentiable relation, and a
fixed receding-source video supplies the same visual target to guided and
unguided audio resampling. The experiment asks whether correcting that
relation improves physical behavior without reducing the test to the
optimized score alone.

\subsection{Guidance Objective and Calibration}
\label{supp:guidance_objective}

The intervention holds the canonical video fixed and resamples only the
audio stream. Here, model refinement denotes improving the model's
inference-time generation process through diagnostic-guided sampling;
no parameter update or retraining is involved. The same mechanism-level
diagnoses can also inform future training objectives, reward models, and
candidate-selection strategies. The intervention does not modify the
prompt, regenerate the video, or change the Range Attenuation evaluator.
Let \(d_i\) denote the relative visual range at time \(i\)
and \(e_i\) the differentiable envelope extracted from the decoded
mel-like representation. We optimize
\begin{equation}
  \mathcal{L}_{\mathrm{dist}}
  =
  \sum_{i,j}
  \left[
    \log\frac{e_i+\epsilon}{e_j+\epsilon}
    -
    \gamma\log\frac{d_j+\epsilon}{d_i+\epsilon}
  \right]^2.
  \label{eq:distance_guidance}
\end{equation}
Here \(\gamma\) specifies the target attenuation exponent; it is not the
loss weight. We use \(\gamma=1\), corresponding to inverse-distance
pressure or amplitude attenuation.

\paragraph{Decoded-mel implementation}
At each denoising step, we unpatchify the audio \(x_0\) latent using an
AudioPatchifier with patch size one. The default latent has eight
channels and 16 mel bins. If \(z_{c,m,t}\) denotes the Audio VAE
decoder output, the differentiable envelope is
\begin{equation}
  \begin{aligned}
    x_{c,m,t}
      &= \exp\!\left(\operatorname{clamp}(z_{c,m,t},-20,20)\right),\\
    e_t
      &= \sqrt{\operatorname{mean}_{c,m}\!\left[x_{c,m,t}^{2}\right]
          + \epsilon},
      \qquad \epsilon=10^{-6}.
  \end{aligned}
  \label{eq:decoded_mel_envelope}
\end{equation}
We interpolate the canonical visual-distance trajectory to this decoded
mel time axis. All upper-triangular valid token pairs \((i,j)\) are
eligible; both endpoints must satisfy \(e_i,e_j>10^{-5}\). When more
than 4,096 pairs are available, deterministic evenly spaced indices
select 4,096 pairs rather than randomly sampling pairs.

Equation~\ref{eq:distance_guidance} presents the squared-residual form
of the objective. The implementation applies Smooth L1 loss with Huber
transition \(\beta=0.05\) to the same log-ratio residual, reducing the
effect of isolated envelope tokens. We differentiate this decoded-mel
loss with respect to audio \(x_0\), update in the negative-gradient
direction with guidance scale 5, and clip the update norm to
\(0.01\lVert x_0\rVert\).

\paragraph{Decoded-mel proxy calibration}
We randomly sample 40 clean, non-floor audio clips from a calibration
pool disjoint from the 126 guidance-evaluation samples. The non-floor
criterion excludes clips whose decoded envelopes reach the numerical
floor required by the loss. For each retained clip, we synthesize distances
\(r\in\{1,1.5,2,3,4\}\), scale the waveform by \(r_0/r\), and pass every
version through the same preprocessing, Audio VAE encoder, Audio VAE
decoder, decoded-mel conversion, and envelope extraction used during
guidance. We fit
\begin{equation}
  \log e_{n,r}
  =
  \alpha_n-\hat{\gamma}\log r+\varepsilon_{n,r},
  \label{eq:proxy_calibration}
\end{equation}
where \(\alpha_n\) is a clip-specific intercept.

\begin{table}[!ht]
  \centering
  \begin{tabular}{lr}
    \toprule
    Calibration statistic & Value \\
    \midrule
    Held-out clips & 40 \\
    Distance points & 200 \\
    \(\hat{\gamma}\) & 0.9971 \\
    Bootstrap 95\% CI & [0.9914, 1.0014] \\
    Fixed-effect \(R^2\) & 0.9997 \\
    Per-clip exponent mean & 0.9971 \\
    Per-clip exponent std. & 0.0165 \\
    Per-clip exponent median & 1.0013 \\
    Per-clip exponent range & [0.9182, 1.0112] \\
    Clamp saturation rate & 0 \\
    Numerical-floor rate & 0 \\
    \bottomrule
  \end{tabular}
  \caption{Calibration of the decoded-mel envelope under injected
  \(1/r\) waveform attenuation.}
  \label{tab:proxy_calibration}
\end{table}

The fitted exponent is nearly one, supporting the interpretation of the
decoded-mel envelope as an amplitude-like proxy and motivating
\(\gamma=1\) in Equation~\ref{eq:distance_guidance}.

\subsection{Sampling Configuration}
\label{supp:sampling_configuration}

We compare three conditions: the original canonical output, an audio
resample without guidance, and an audio resample with distance guidance.
All three conditions use exactly the same canonical LTX-2.3
receding-source video and visual range trajectory. The no-guidance and
guided resamples additionally use the same prompt and random seed,
\(270626+\text{selected rank}\), so their audio differences isolate the
guidance update rather than resampling randomness. Each successful
sample uses eight denoising steps, and the guidance objective is
scheduled at every step; the envelope and pair-validity gates determine
whether a step produces an actual update. After filtering failed mux or
audio-extraction cases, each condition contains 126 valid paired outputs.

\begin{table}[!ht]
  \centering
  \begin{tabular}{lr}
    \toprule
    Parameter & Setting \\
    \midrule
    Target exponent \(\gamma\) & 1 \\
    Guidance scale & 5 \\
    Maximum update-norm ratio & 0.01 \\
    Start fraction & 0.0 \\
    Update frequency & Every step \\
    Maximum token pairs & 4,096 \\
    Huber transition \(\beta\) & 0.05 \\
    Envelope \(\epsilon\) & \(10^{-6}\) \\
    Envelope RMS gate & \(10^{-5}\) \\
    Denoising steps & 8 \\
    Distance-ratio clipping & None \\
    Gain-ratio cap & None \\
    \bottomrule
  \end{tabular}
  \caption{Distance-guidance sampling configuration.}
  \label{tab:guidance_configuration}
\end{table}

\begin{figure*}[!t]
  \centering
  \includegraphics[width=\textwidth,height=0.64\textheight,keepaspectratio]
    {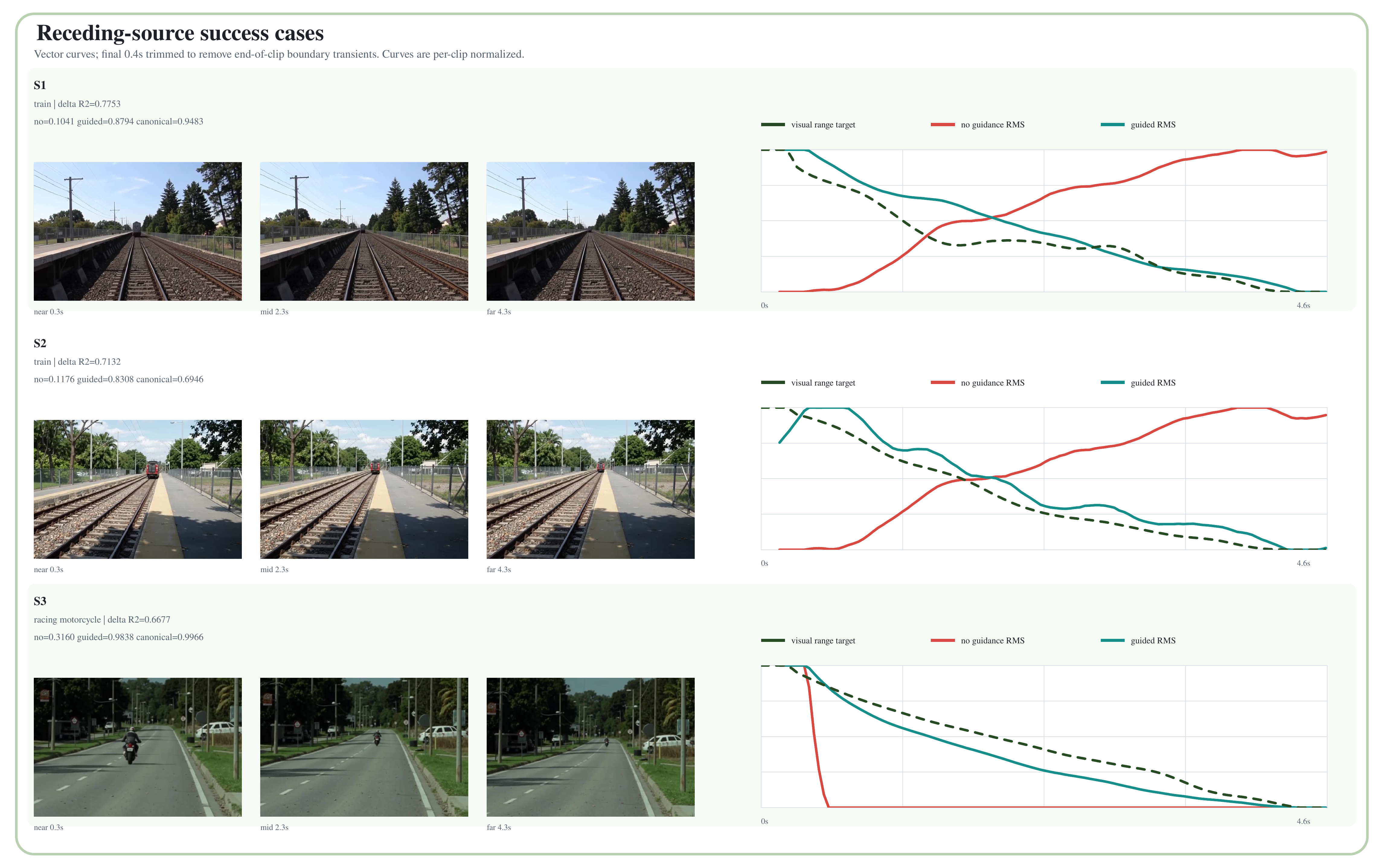}
  \caption{Three receding-source success cases selected by paired
  \(R^2\) improvement. Each row shows fixed-video key frames and
  within-clip normalized visual-range, no-guidance RMS, and guided RMS
  trajectories.}
  \label{fig:guidance_success_cases}
\end{figure*}

\begin{table*}[!t]
  \centering
  \small
  \begin{tabular*}{\textwidth}{@{\extracolsep{\fill}}lrrrrlrr@{}}
    \toprule
    Metric & Canonical & No guidance & Guided & Paired \(\Delta\) &
      95\% CI & \(p\) & Win rate \\
    \midrule
    Mean \(R^2\) & 0.8163 & 0.7138 & 0.8580 & 0.1442 &
      [0.0996, 0.1889] & \(<10^{-4}\) & 80.16\% \\
    Spearman \(\rho\) & \(-0.3411\) & \(-0.0880\) & \(-0.4356\) & \(-0.3476\) &
      [-0.4441, -0.2446] & \(<10^{-4}\) & 80.6\% \\
    CLAP & 0.1029 & 0.0498 & 0.1049 & 0.0551 &
      [0.0287, 0.0823] & \(<10^{-4}\) & 61.2\% \\
    Integrated LUFS & \(-23.601\) & \(-23.0388\) & \(-23.5864\) & \(-0.5476\) &
      [-2.4660, 1.4612] & 0.5923 & -- \\
    Audiobox PQ & 6.031 & 6.106 & 5.884 & -- & -- & -- & -- \\
    Clipping sample ratio & 0.0000 & 0.0000 & 0.0000 & 0.0000 &
      [-0.0000, 0.0001] & 0.1753 & 2.9\% \\
    \bottomrule
  \end{tabular*}
  \caption{Target attenuation and non-target quality metrics under
  fixed-video audio resampling. Where reported, paired differences are
  guided minus no guidance; intervals and two-sided sign-flip tests are
  computed on paired outputs.}
  \label{tab:guidance_results}
\end{table*}

\subsection{Target and Non-Target Metrics}
\label{supp:guidance_ablations}

The target improvement is interpreted together with independent CLAP,
integrated-LUFS, and Audiobox PQ measurements rather than from the
optimized score alone.

Unguided audio resampling reduces mean Range Attenuation \(R^2\) from
0.8163 to 0.7138. Guidance increases it to 0.8580 and outperforms the
unguided resample in 80.16\% of cases. The paired \(R^2\) improvement is
0.1442 [0.0996, 0.1889], with \(p<10^{-4}\). The mean Spearman
correlation shifts from \(-0.0880\) to \(-0.4356\), a paired change of
\(-0.3476\), indicating a stronger inverse relation between visual
range and audio level.

CLAP similarity increases by 0.0551 [0.0287, 0.0823]. In contrast, the
integrated-LUFS interval includes zero, supporting that the intervention
is not merely a global loudness adjustment. Audiobox PQ decreases from
6.106 without guidance to 5.884 with guidance, while remaining close to
the canonical value of 6.031. We therefore interpret the result as a
targeted physical improvement with a modest predicted-quality tradeoff
rather than no degradation or an across-the-board quality gain.

This single-model, single-relation intervention is an actionability
proof of concept.

\end{document}